\documentclass[useAMS,usenatbib]{mn2e}
\usepackage{epsfig}
\title[Flux and colour variations in blazars]{Short Term Flux and Colour Variations in Low-Energy Peaked Blazars} 

\author[Rani et al.]
{Bindu Rani$^{1}$\thanks{E-mail: bindu@aries.res.in}, 
Alok C.\ Gupta$^{1}$, 
A. Strigachev$^{2}$, 
R. Bachev$^{2}$, 
Paul J. Wiita$^{3,4}$, 
\newauthor  E. Semkov$^{2}$,
E. Ovcharov$^{5}$, 
B. Mihov$^{2}$, 
S. Boeva$^{2}$, 
S. Peneva$^{2}$, 
B. Spassov$^{2}$, 
\newauthor  S. Tsvetkova$^{2}$, 
K. Stoyanov$^{2}$, 
A. Valcheva$^{5}$ 
\\ 
$^{1}$Aryabhatta Research Institute of Observational Sciences (ARIES),
Manora Peak, Nainital -- 263129, India\\
$^{2}$Institute of Astronomy, Bulgarian Academy of Sciences, 72 Tsarigradsko
Shosse Blvd., 1784 Sofia, Bulgaria\\
$^{3}$Department of Physics and Astronomy, Georgia State University, P.O.\ Box 4106, Atlanta,
GA 30302--4106, USA\\
$^{4}$Department of Physics, The College of New Jersey, P.O.\ Box 7718, Ewing, NJ 08628, USA\\
$^{5}$Department of Astronomy, University of Sofia, 5 James Bourchier, 1164 Sofia, Bulgaria
}

\begin{document}

\date{Accepted ....... Received  ......; in original form ......}

\pagerange{\pageref{firstpage}--\pageref{lastpage}} \pubyear{2010}

\maketitle

\label{firstpage}

\begin{abstract}
We have measured multi-band optical flux and colour variations for a sample
of 12 low energy peaked blazars on short, day-to-month, timescales.  Our sample contains six BL Lacertae
objects and six flat spectrum radio quasars. These  photometric observations, made during
 September 2008 to June 2009, used five optical telescopes, one in India and four in Bulgaria.
We detected short term flux variations in eleven of these blazars
and colour variability in eight of them.
Our data indicate that six blazars (3C 66A, AO 0235$+$164,
S5 0716$+$714, PKS 0735$+$178, OJ 287 and 3C 454.3) were observed in pre- or post-outburst states,
that five (PKS 0420$-$014, 4C 29.45, 3C 279, PKS 1510$-$089 and BL Lac) were in a low state, while
one (3C 273)  was in an essentially steady state.  The duty cycles for flux and colour variations on short timescales
in these low energy peaked blazars are $\sim$92 per cent and $\sim$ 33 per cent, respectively.
The colour vs magnitude correlations seen here support the hypothesis that BL Lac objects tend to become bluer with
increase in brightness; however,  flat spectrum radio quasars may show the opposite trend, and there
are  exceptions to these trends in both categories of blazar.  We briefly discuss
 emission models for active galactic nuclei that might explain our results.
\end{abstract}


\section{Introduction}

BL Lacertae objects (BL Lacs) and flat-spectrum radio quasars (FSRQs)  are now usually clubbed together and called blazars; they represent
a small subset of radio-loud active galactic nuclei (AGNs) characterized by strong and rapid flux
variability across the entire electromagnetic spectrum. Blazars exhibit strong polarization from radio
to optical wavelengths and usually have core-dominated radio structures. These extreme AGNs  provide a natural 
laboratory to study the mechanisms of energy extraction from the central supermassive black holes
and the physical properties of jets and perhaps accretion disks. According to the orientation based unified model 
of radio-loud AGNs, blazar jets usually make an angle $\leq 10^{\circ}$ to our line-of-sight (e.g., Urry \& 
Padovani 1995).  \par
The electromagnetic (EM) radiation from blazars is predominantly 
non-thermal. At lower frequencies (through 
the UV or X-ray bands) the mechanism is almost certainly synchrotron emission while at higher frequencies the 
emission mechanism is probably due to Inverse-Compton (IC) emission (e.g., Sikora \& Madejski 2001; Krawczynski 2004).
The spectral energy distributions (SEDs) of blazars have a double-peaked structure (e.g., Giommi, Ansari \& Micol 1995; 
Ghisellini et al.\ 1997; Fossati et al.\ 1998). Based on the location of the first peak of their SEDs, $\nu_{\rm peak}$, 
and blazars are often sub-classified into 
 low energy peaked blazars (LBLs) and high energy peaked blazars (HBLs)  (Padovani \& Giommi 1995); 
however, it should be noted that the SED peaks can be located at intermediate frequencies as 
well, giving rise to the intermediate peaks blazar (IBL) classification (e.g.,  Sambruna, Maraschi \& Urry 1996). 
The first component 
peaks in the near-infrared (NIR)/optical in case of LBLs and at UV/X-rays in HBLs, while the second component usually 
peaks at GeV energies in LBLs and at TeV energies in HBLs.  More specifically, Nieppola, Tornikoski \& Valtaoja (2006)
classify over 300 BL Lacs and suggest that blazars with $\nu_{\rm peak} \approx 10^{13-14}$ Hz are  LBLs, those with
$\nu_{\rm peak} \approx 10^{15-16}$ Hz are IBLs and those with $\nu_{\rm peak} \approx 10^{17-18}$ Hz are HBLs.\par
Observations of blazars reveal that they are variable at all accessible timescales, from a few tens of minutes to years and even decades at many frequencies (e.g., Aller et al.\ 1992; Carini \& Miller 1992a; Carini et al.\ 1992b; Wagner \& Witzel 1995; Gupta et al.\ 2004; Ter{\"a}sranta et al.\ 2004;
Ostorero et al.\ 2006; Villata et al.\ 2007; D'Ammando et al.\ 2009). Based on these different
timescales, the variability of blazars can be broadly divided into three classes, e.g., intra-day variability
(IDV), short-term variability (STV), and long-term variability (LTV). Variations in the flux of 
a source of a couple of hundredths of a magnitude up
to a few tenths of magnitude over a time scale of a day or less is known as IDV
(Wagner \& Witzel 1995) or microvariability or intra-night optical variability. 
Flux changes over days to a few 
months are often considered to be STV, while those taking from several months to many years are usually called LTV
(e.g., Gupta et al.\ 2004); both of these classes of variations typically exceed $\sim$ 1 magnitude and can 
exceed even 5 magnitudes. \par
The study of variability is one of the most powerful tools for revealing the nature of blazars and
probing the various processes occurring in them.  Over the last two decades, the optical/NIR variability
of blazars has been extensively studied on diverse timescales (e.g.,  Miller et al.\ 1989; Carini et al.\ 1992b;
Courvoisier et al. 1995; Heidt \& Wagner 1996; Takalo et al. 1996; Sillanp{\"a}{\"a} et
al. 1996a, 1996b; Bai et al.\ 1998, 1999; Fan et al.\ 1998; Xie, et al.\,  2002a; Stalin et al.\ 2005, 2006;
Gupta et al.\ 2004, 2008a, 2008b, 2008c, 2008d, 2009a; Ciprini et al.\ 2003, 2007; Ter{\"a}sranta et al.\ 1992, 2004; Rani et al.\ 2009; and references therein).
There are several theoretical models that might be able to explain the observed variability over wide time-scales 
for all bands, with the leading contenders based upon shocks propagating down relativistic jets (e.g., Marscher \& Gear 
1985; Qian et al.\ 1991; Hughes, Aller \& Aller 1991; Marscher, Gear \& Travis 1992; Wagner \& Witzel 1995; Marscher 
1996). Some of the variability may arise from helical structures, precession or other geometrical effects occurring within 
the jets (e.g., Camenzind \& Krockenberger 1992; Gopal-Krishna \& Wiita 1992) and some of the radio variability is due 
to extrinsic propagative effects (e.g., Rickett et al.\ 2001).  Hot spots or disturbances in or above accretion  
disks surrounding the black holes at the centers of AGNs  (e.g., Mangalam \& Wiita 1993; Chakrabarti \& Wiita 1993) are 
likely to play a key role in the variability of non-blazar AGNs and might provide seed fluctuations that could be advected 
into the a blazar jet and then Doppler amplified. \par    

In this work we aimed to search for STV in the multi-band fluxes and colours of a sample of twelve
LBLs that consists of six BL Lacs and six FSRQs and to thereby cast more light on  the emission mechanisms responsible for these variations.  Our work is mainly focused on day to 
day variations in the magnitudes as well as the colours of these dozen sources which were the brightest blazars 
visible from India and Bulgaria.
Throughout this paper when we use the term colour variation we mean a change in the colour of a blazar with time.
Optical flux variations in blazars are often 
associated with changes in their spectral
shapes.  These can be quantified by studying the properties of the colour indices of each  source. So we have also
investigated the spectral variabilities in the observed visible wavebands of LBLs and have investigated if there are differences between  
spectral variability trends in those classified as BL Lacs and those called FSRQs.  We have also studied the spectral 
variability of these LBLs with changes in the brightness of the sources. \par
The paper is structured as follows.  In Section 2, we present the observations and data reduction 
procedure.  Section 3 provides our results, which are discussed in  Section 4.  We present our conclusions 
in Section 5.

\section{Observations and Data Reduction}

Our observations of twelve LBLs were performed using five optical telescopes, 
one in India and four in Bulgaria.  All of these telescopes are equipped with CCD detectors and Johnson UBV and Cousins RI filters. Details about the telescopes, detectors and other parameters related to the observations
are given in Table 1.  A complete log of observations of these LBLs from these five telescopes is given in
Table 2.   \par
We carried out optical photometric observations  during the period September
2008 to June 2009.  The raw photometric data  were processed by standard methods which are briefly described
here.  For image processing or pre-processing, we generated a master bias frame for each observing night
by taking the median of all bias frames. The master bias frame for the night was subtracted from all flat and
source image frames taken on that night. Then the master flat in each passband was generated by median
combine of all flat frames in each passband. Next, the normalized master flat for each passband was generated.
Each source image frame was flat-fielded by dividing by the normalized master flat in the respective band to 
remove pixel-to-pixel inhomogeneities. Finally, cosmic ray removal was done from all source image
frames. Data pre-processing used the standard routines in Image Reduction and Analysis
Facility\footnote{IRAF is distributed by the National Optical Astronomy Observatories, which are operated
by the Association of Universities for Research in Astronomy, Inc., under cooperative agreement with the
National Science Foundation.} (IRAF) and ESO MIDAS software.  \par
We processed the data using the Dominion Astronomical Observatory Photometry (DAOPHOT II) 
software to perform the circular concentric aperture photometric technique (Stetson 1987, 1992). For each night we carried 
out aperture photometry with four different aperture radii, i.e., 1$\times$FWHM, 2$\times$FWHM, 
3$\times$FWHM and 4$\times$FWHM. On comparing the photometric results we found that 
aperture radii of 2$\times$FWHM almost always provided the best S/N ratio, so we adopted that aperture for our 
final results.   \par
For each of the twelve blazars, we observed more than three local standard stars on the 
same field. The magnitudes of the standard stars that we measured in the fields of our target sources are 
given in Table 3.  Having multiple standard stars was  required to confirm that the final standard stars were
truly  non-variable.  We employed two non-varying standard stars from each blazar field with magnitudes similar to those of the target and plotted 
their differential instrumental magnitudes so as to obtain an indication of the noise in the measurements. 
Finally, to calibrate the photometry of the blazars, we used the one standard star which had a colour closer to 
that of the blazar. The calibrated light curves (LCs) of these blazars are plotted in the same panel with the differential 
instrumental magnitudes of those two standard stars in the figures  1 $-$ 12.  We used  R\footnote{R: A language and environment for statistical computing. 
R Foundation for Statistical Computing, Vienna, Austria. ISBN 3-900051-07-0, URL http://www.R-project.org.} to write additional programs for data 
processing.

\section[]{Results}

\subsection{Variability  Parameters}

To search for and describe blazar variability we have employed  two quantities commonly used in 
the literature, the variability detection parameter, $C$, and the variability amplitude, $A$. The former was introduced by Romero et al. (1999), and is defined as the average of $C1$ and $C2$ where
\begin{equation}
C1 = \frac{\sigma(BL - starA)}{\sigma(starA - starB)} \hspace{0.2 cm} {\rm \&} \hspace{0.2 cm} C2 = \frac{\sigma(BL - starB)}{\sigma(starA - starB)}.
\end{equation}
Here $(BL - starA)$ and $(BL - starB)$ are the differential instrumental magnitudes of the blazar and standard star A and the blazar
and standard star B, respectively, while $\sigma (BL-starA)$, $\sigma (BL-starB)$ and $\sigma (starA-starB)$ are the observational 
scatter of the differential instrumental magnitudes of the blazar and star A, the blazar and star B and starA and star B,
respectively. If $C \geq 2.57$, the confidence level
of a variability detection is $> 99$\%, and we consider this to be a positive detection of a variation. 
For a few sources, the (star A - star B) LCs show isolated apparently discrepant points, that we conservatively retain in our analysis.  
These points increase the $\sigma(starA - starB)$ values relatively more than the BL--star scatter, thereby slightly reducing the $C$ values and the confidence with which variability claims are made.
\par
Heidt \& Wagner (1996) introduced the variability amplitude, defined as
\begin{equation}
A = 100\times \sqrt{{(A_{max}-A_{min}})^2 - 2\sigma^2}(\%),
\end{equation}  
where $A_{max}$ and $A_{min}$ are the maximum and minimum magnitudes in the calibrated LCs of
the blazar and the average measurement error of the blazar LC is $\sigma$ .
\subsection{Short-Term Flux and Colour Variability of Individual Blazars} 

\noindent
The LCs in the BVRI filters and the time curves of the B$-$V, V$-$R, R$-$I and B$-$I colours 
are displayed in Figures 1$-$12, one figure for each source, plotted along with the corresponding 
curves of the  standard stars used for comparison. The complete observing log for the blazars in given 
in Table 2. Table 3 contains the standard stars observed in the fields of the blazars.  The estimated $A$ and $C$  values of the individual blazars are listed in Table 4. 
We now report some key individual results for each of our sources, placed in context of earlier work.

{\bf 3C 66A:} This BL Lac object is a low energy peaked blazar (LBL); although its redshift
is usually taken as $z = 0.444$ (Lanzetta et al.\ 1993) that value is actually quite uncertain
(e.g., Bramel et al.\ 2005).
 Since its optical identification (Wills \& Wills 1974), the source has been regularly monitored at 
optical frequencies, and somewhat less regularly at radio frequencies (Aller et al.\ 1992; Takalo et al.\ 1996). Fan \& Lin
(1999, 2000) have studied the long-term optical and IR variability of the source and reported a variation of
$\la$1.5 mag at time scales of $\sim$1 week to several years at those two frequencies. B{\"o}ttcher
et al.\ (2005) reported a rapid microvariability of $\sim$0.2 mag within 6 hours and they also reported several 
major outbursts in the source separated by $\sim$50 days and argued that the outbursts seem to have
quasi-periodic behaviour.  Toward  the end of 2007 3C 66A was found to be in an optically active phase, which triggered a new
optical-IR-radio Whole Earth Blazar Telescope (WEBT) campaign on the source (B{\"o}ttcher et al.\ 2009a). \par
The source 3C 66A showed large flux, but no significant colour, variations during our observing run (Fig.\ 1). The 
source first faded by $\sim$0.6 magnitude within
10 days and then it brightened by $\sim$0.9 magnitude within 80 days in all
four of the observed bands. During the observing run the maximum brightness detected in
the R band was 14.0 magnitudes, which is only $\sim$0.6 magnitudes fainter than the brightest 
(R $\simeq$13.4) reported in the source by B{\"o}ttcher et al. (2009a).
We may have observed the source in a pre-outburst state since the light curve showed an overall brightening trend. 

\noindent
{\bf AO 0235$+$164:} {\bf This} blazar was classified as a BL Lac object
by Spinrad \& Smith (1975).  AO 0235$+$164 has been seen to be highly variable over
all timescales and at all frequencies (e.g., Ghosh et al.\ 1995; Heidt et al.\ 1996; Raiteri et al.\ 2001) and a very high 
fractional polarization of $\sim$40$\%$ has been reported in the source both 
at visible and IR frequencies (e.g., Impey et al.\ 1982). By analysing 25 years (1975$-$2000) of optical and radio
data, Raiteri et al.\ (2001) argued that the source seemed to have an optical outburst period of $\sim$5.70 years
but the expected outburst in 2004 was not detected by a 2003--2005 multiwavelength WEBT observing campaign (Raiteri
et al.\ 2006a).  A more detailed long term optical data analysis suggested a possible outburst period of $\sim$8 years
in this source (Raiteri et al.\ 2006b) and this period was supported by observations of
Gupta et al. (2008c).  Strong
IDV flux variations of 9.5$\%$ and 13.7$\%$ during two nights were observed by Gupta et al. (2008c).
Rani et al.\ (2009) recently reported a high probability of the presence of nearly periodic fluctuations with a timescale 
of $\sim$17.8 days in a 12 year long X-ray light curve obtained by the  the Rossi X-ray Timing Explorer 
 satellite's All Sky Monitor  instrument.  \par

During our observations AO 0235$+$164 showed significant flux as well as colour variations (Fig.\ 2).  The source
first brightened by $\sim$0.9 magnitude within an interval of 46 days and then it faded by $>$2 mag within
94 days in all four bands. An outburst state in Jan 2007 was reported by
Gupta et al.\ (2008c) with R$_{mag}$ $\sim$14.98 being the peak during their observing run. In our observation span the
source reaches a maximum brightness of R = 15.18 mag and then fades to R = 17.4 mag, so it seems likely that we have
caught  this BL Lac in a post-outburst phase.  \par

\noindent
{\bf PKS 0420$-$014:} The blazar PKS 0420$-$014 is classified as a FSRQ. It has
been observed in optical bands for four decades.  Several papers have reported multiple optically active and
bright phases of the source and perhaps regular major flaring cycles (e.g., Villata et al.\ 1997;  Webb et al.\ 1998;
Raiteri et al.\ 1998a and references therein). An increase
of $\sim$ 2$-$3 magnitudes during the active phases of this blazar was reported by Webb et al.\ (1998)  during  observations that stretched
from December 1969 to January 1986.  Variations of 2.8 mag with a time
scale of $\sim$22 years have been reported (Clements et al.\ 1995). Two modes of variability at radio wavelengths superimposed on a $\sim$13 months
nearly regular cycle were suggested through 5 years of radio monitoring in the Hamburg Quasar Monitoring
Program (Britzen et al.\ 2000).  \par

During our observations PKS 0420$-$014  exhibited significant flux and colour variations (Fig.\ 3). Raiteri et al.\ (1998) reported that the historical peak in the light curve
of this source reached R = 14.15 mag after which it decayed to 16.9 mag. The light curve of the source during
our observing run  varies from 16.9 to 16.3 mag in R  with an average of $\sim$16.6 mag, which
is $\sim$2.5 mag fainter than brightest reported value for this source. Since our values are close to
 the faintest reported for this source, we conclude that  PKS 0420$-$014 was probably in a low state in the period we observed it.  \par

\noindent
{\bf S5 0716$+$714:} This blazar is classified as a BL Lac object. It is important to note that
this source has been classified 
as an IBL by Giommi et al.\ (1999) since the frequency of the first SED peak varies
between 10$^{14}$ and 10$^{15}$ Hz, and thus does not fall into the
wavebands specified by the usual definitions of LBLs and HBLs.  More
recently, however, Nieppola et al.\ (2006) studied the SED distribution of a large sample of BL Lac objects and categorized 0716$+$714  as a LBL; we adopt that classification in this paper. The optical continuum of the source is so
featureless that it was hard to estimate its redshift but there is a very recent claim of
 $z = 0.31\pm0.08$ by Nilsson et al.\ (2008). This source has been
extensively studied at all observable wavelengths from radio to $\gamma$-rays on diverse time scales
(e.g., Wagner et al.\ 1990; Heidt \& Wagner 1996;  Giommi et al.\ 1999; Villata et al.\ 2000; Raiteri et al.\ 2003; Montagni et al.\
2006; Foschini et al.\ 2006; Ostorero et al.\ 2006; Gupta et al.\ 2008a,
2008c and references therein). This source is one of the brightest BL Lacs in optical bands and has an IDV duty
cycle of nearly 1.  Unsurprisingly, S5 0716$+$714 has been the subject of several optical monitoring campaigns on IDV timescales (e.g., Wagner
et al.\ 1996; Sagar et al.\ 1999; Montagni et al.\ 2006; Gupta et al.\ 2008c, and references therein). This
source has shown five major optical outbursts (Gupta et al.\ 2008c) separated by $\sim$3.0$\pm$0.3 years.
High optical polarizations of $\sim$ 20$\%$ -- 29$\%$ have also been observed in the source (Takalo et al.\ 1994;
Fan et al.\ 1997). Recently, Gupta, Srivastava \& Wiita (2009) analysed the excellent  intraday optical LCs of the source obtained by Montagni et al.\ (2006) and reported good evidence for nearly periodic oscillations ranging between 25 and 73 minutes on several different nights. \par

Our LCs of S5 0716$+$714 showed very significant variations of $\sim$2 mag in each of the observed bands; however, no 
significant colour variation is observed in the source except for B$-$I (Fig.\ 4). A large increase in the brightness of the source occurred 
between the two early observations and the 
seven later ones: over 230 days the source brightened by 2.2 magnitudes (15.4 to 13.2) in  R, becoming 
only $\sim$0.65 magnitude fainter than the brightest magnitude (R = 12.55 mag) reported for the source (Gupta et al.\ 2008c). 
Assuming that the source continued its brightening following a linear trend, it should have
 reached 
R$=$12.55 mag in September 2009.  The calculated time difference between the last known
outburst  in January 2007 and this possible outburst is $\sim$2.7 years. This temporal gap  
is consistent with the optical outburst time 3.0$\pm$0.3 years in the long term optical period that was
earlier reported for this source (Gupta et al.\ 2008c).  Therefore we organized a radio-optical observing campaign for five days in December 2009 to ascertain the source behaviour and state; the analysis of this data is underway.
 \par

\noindent
{\bf PKS 0735$+$178:} This blazar  has been classified as a BL Lac object (Carswell et al.\ 1974).
There have been several papers concerning  its redshift determination 
(e.g., Carswell et al.\ 1974; Falomo \& Ulrich 2000, and references therein) 
with the most recent result of  $z = 0.424$  for PKS 0735$+$718 found  using a HST snapshot image (Sbarufatti et al.\ 2005).
Since it was optically identified, this source has been 
extensively observed across the whole EM spectrum (Ter{\"a}sranta et al.\ 2004; Fan \&
Lin 2000; Gu et al.\ 2006; Gupta et al.\ 2008c; Ciprini et al.\ 2007 and references therein.).  Radio frequency observations show slow variability with some outbursts (Ter{\"a}sranta et al.\ 1992, 2004),
but hardly any correlation has been found between the radio and optical flares (Clements et al.\ 1995; Hanski 
et al.\ 2002; Ciaramella et al.\ 2004). A periodicity of $\sim$14 years has been suggested to be present in the source
using a century long optical light curve (Fan et al.\ 1997).  Optical variability on 
IDV and STV timescales has naturally been observed for 0735$+$178 (Xie et al.\ 1992; Massaro et al.\ 1995; Fan et al.\ 1997; 
Zhang et al.\ 2004; Ciprini et al.\ 2007; Gupta et al.\ 2008c).  Significant fractional polarizations  ($\sim$ 1$\%$ to 30$\%$) 
have been observed  both at optical and IR bands (Mead et al.\ 1990; Takalo et al.\ 1991, 1992b; Valtaoja et al.\ 
1991a, 1993; Tommasi et al.\ 2001).   \par

We found strong flux variations in all the well observed passbands for PKS 0735$+$178 (Fig.\ 5);
however, as there are only two 
data points in the B band LC, we cannot discuss the nature of its variation. Except for R$-$I, 
no significant colour variation was seen in the source. 
The average R$_{mag}$ of PKS 0735$+$178 during our observing run was 15.80 which is $\sim$1.8 mag
fainter than the brightest  (R $\sim$ 14.0 mag) and $\sim$1.2 mag brighter than the faintest
magnitude (R$_{mag}$ $\sim$ 17.0) reported in the source (Ciprini et al.\ 2007). 
Thus we have probably observed the source in either a pre- or post-outburst state as long as
 there have
been no long-term changes in the underlying light-curve.  \par

\noindent
{\bf OJ 287:} This BL Lac object is one of the most extensively observed  for variability; 
it is also among the very few AGN's whose substantial brightness means
that more than a century of optical observations are available (e.g., Carini et al.\ 1992b;
Sillanp{\"a}{\"a} et al.\ 1996a, 1996b; Fan et al.\ 2002, 2009; Abraham et al.\ 2000;
Gupta et al.\ 2008c). Using the binary black hole
model (Sillanp{\"a}{\"a} et al.\ 1988) for the long-term optical light curve of the source, an outburst
with a predicted  $\sim$12 year period
was detected in the source by the OJ-94 programme (Sillanp{\"a}{\"a} et al. 1996a;
Valtonen et al.\ 2008). On re-analysing the optical data from OJ-94 project, Wu et al.\ (2006) reported
another possible timescale, suggesting a periodicity of $\sim$40 days. Very high optical polarization that is  variable in both
degree and angle   has been  reported in the source (Efimov et al.\ 2002).
The observational properties of OJ 287 from radio to X-ray energy bands have been reviewed by Takalo
et al.\ (1994). During their observations spanning 
2002 to 2007 Fan et al.\ (2009), reported large  variations in the source of
$\Delta$V = 1.96 mag, $\Delta$R = 2.36 mag, and $\Delta$I = 1.95 mag.  \par

We found OJ 287 to have significant flux variations in all the observed passbands but 
no significant colour variation was found (Fig.\ 6).   A flare in the LCs of the blazar OJ 287 of $\sim$1 mag was observed in all bands over an interval
 of 36 days, but we had only one night of data near the flare's peak so it is quite likely  that the maximum of the flare
was even stronger.  It is interesting that this result is consistent with a $\sim$37 day period of rotation of the the plane
of polarization seen of OJ 287 (Efimov et al.\ 2002) and close to a 40 day periodicity reported in source by
Wu et al.\ (2006). 
The average R band magnitude (R $\sim$ 14.7 mag) of the source during our observing run is $\sim$2.5 mag
fainter than the brightest (R = 12.09 mag) and $\sim$2.3 magnitude brighter than the faintest magnitude
(R = 16.47 mag) reported in the source by Fan et al.\ (2000).  So it seems that we observed OJ 287
in an intermediate state.  \par

\noindent
{\bf 4C 29.45:} This optically violent variable quasar  belongs to the class of FSRQs.
At both optical and IR bands STV and LTV have been observed in this source (Branly et al.\ 1996;
Noble \& Miller 1996; Ghosh et al.\ 2000). A very large variation
($\Delta$V $>$ 5 mag) has been reported in the source during the optical outburst of 1981
(Wills et al.\ 1983). Large fractional polarizations, up to $\sim$28$\%$, have been observed in 4C 29.45 both at optical and IR frequencies (Holmes et al.\ 1984; Mead et al.\ 1990). Its optical flux and colour variations  have been recently studied by Fan et al.\ (2006). They have reported
amplitude variations of $\sim$4.5 -- 6 mag in all passbands (U, B, V, R, I) and also found that
there were possible periods of 3.55 or 1.58 years in the long term optical light curve of the 
source.  \par

Our LCs of 4C 29.45 indicate that it is variable in all the observed bands but the percentage of variation differs substantially
among them, leading to the significant colour variations (Fig.\ 7). During our observing run 4C 29.45 varied from 
16.30 to 17.57 magnitude in the I band.  The faintest
magnitude we observed in the source is thus 0.12 magnitudes fainter that
the faintest level,  I = 17.45$\pm$0.05 mag, reported for this source by Fan et al.\ (2006) so we
clearly observed the source in a faint phase.  The   faintest state of the source
was on 23 May 2009 but  it had brightened by $\sim$0.2 mag by the next day.  \par

\noindent
{\bf 3C 273:} The FSRQ 3C 273 was the first quasar discovered (Schmidt 1963).
Categorized as a LBL (Nieppola et al.\ 2006), its spectral energy distribution, correlations
among flares in different energy bands and the approaching jet's orientation have been extensively
 studied at all EM bands (e.g., Valtaoja et al.\ 1991b; Takalo et al.\ 1992a, 1992b). Many papers cover 3C 273's 
observational properties of the source in the visible band
(Angione et al.\ 1981; Sitko et al.\ 1982; Corso et al.\ 1985, 1986; Moles et al. 1986; Hamuy \& Maza 1987;
Sillanp{\"a}{\"a} et al. 1991; Valtaoja et al. 1991b; Takalo et al. 1992a,
1992b; Elvis et al.\ 1994; Lichti et al. 1995; Ghosh et al.\ 2000;
 Dai et al.\ 2005).  Analyses of the optical light curve of 3C 273 spanning
over 100 years suggests a LTV timescale of $\sim$13.5 years (Fan, Qian \& Tao 2001). Recently,
the STV and colour index properties of the source were studied by Dai et al.\
(2009). They found a strong  correlation between the colour index and brightness of the source in the sense that the spectrum becomes flatter, or bluer, when the source brightens and steeper, or redder, when it fades.  \par

The blazar 3C 273 was essentially in a steady state during our observing run, with  no significant 
variation greater than $\sim$0.03 mag in any of the observed LCs (Fig.\ 8). Nor were any significant variations  found in the colour LCs of the source. Very recently Dai et al.\ (2009) 
reported that the source also was essentially steady  during their observations, which ran from January 2003 to April 2005,
with an average R band magnitude $\simeq$12.44 which was only 0.12 mag brighter than that of our observing run
(R = 12.56 mag).  Thus it is very possible that the same dull state continued in 3C 273 from 2003 through our observations in 2009.  \par

\noindent
{\bf 3C 279:} This FSRQ   and shows strong optical polarization and
flux variabilities at all frequencies. A very large amplitude variation of $\Delta$B $\geq$6.7 in the optical flux of 3C 279 was reported long ago by Eachus \& Liller (1975). A rapid variation of 1.17 mag
within  40 minutes was seen in the V band (Xie et al.\ 1999) and a STV of 0.91 mag in the
 R band was seen over 49 days (Xie et al.\ 2002b). Recently, Gupta et al.\ (2008c)
reported a 1.5 mag variation in the R band of 3C 279 over 42 days but  Webb et al.\ (1990) reported rapid
fluctuations of $\sim$2 mag within 24 hours at visible wavelengths.  Not surprisingly, this source also has been intensively
studied through multi-wavelength campaigns  (Hartman et al.\ 1996; Wehrle et al. 1998). The most
recent WEBT campaign on 3C 279 reported that the source was in the high optical
state and the LCs show quasi-exponential decays of flux of $\sim$1 mag on a time scale of
$\sim$12.8 days (B{\"o}ttcher et al. 2007) which was explained by B{\"o}ttcher \& Principe (2009) as a signature of deceleration of a synchrotron
emitting jet component. \par

It is clear from our LCs that the source is highly variable in all the observed passbands (Fig.\ 9). However, except for  R$-$I. the 
colours of the source are not significantly variable.
3C 279's LC shows a rapid decay in brightness,
i.e., within  5 days $\Delta$V = 1.24 mag, $\Delta$R = 1.10 mag and $\Delta$I = 0.8 mag
which is  faster than most earlier STVs reported in the source (e.g., Xie et al.\ 2002b; B{\"o}ttcher et al.\ 2007; Xie et al.\ 1999; Gupta et al.\ 2008c).
The source was reported to be in outburst in Jan 2007 by Gupta et al.\ (2008c) reaching a brightness of
R $\sim$12.6 mag. The faintest level we observed for 3C 279 was R = 17.1 mag
which is $\sim$4.5 magnitudes fainter than the outburst brightness seen about 2.4 years earlier,
indicating that we caught 3C 279 in a low-state.  \par

\noindent
{\bf PKS 1510$-$089:} This source is classified as a FSRQ and also belongs
to the category of highly polarized quasars. Significant optical flux variations in the source were first
reported by L{\"u} (1972) over a time span of $\sim$5 years. The historical light curve of
PKS 1510$-$089 shows a large variation of $\Delta$B = 5.4 mag during an outburst in 1948 after which it
faded by $\sim$2.2 mag within 9 days (Liller \& Liller 1975).  Strong variations on IDV time scales also have
been reported for PKS 1510$-$089; e.g., $\Delta$R = 0.65 mag within 13 min. (Xie et al.\ 2001), $\Delta$R = 2.0 mag
in 42 min. (Dai et al.\ 2001), $\Delta$V = 1.68 mag in 60 min. (Xie et al.\ 2002a). In the optical LCs of this source, deep minima have been observed  on different days (e.g., Xie et al.\ 2001; Dai et al.\ 2001;
Xie et al. 2002b) that nominally correspond to a time scale of $\sim$42 min, though no more than 3 such
dips were ever seen in a single night so the evidence for a real periodicity is slight.  Nonetheless, an eclipsing binary 
black hole model was proposed to explain the occurrence of these minima (Wu et al.\ 2005a).
This group has also claimed another possible  time scale between minima
of $\sim$89 min (Xie et al.\ 2004). Very recently, while it was being monitored by 
WEBT,  D'Ammando et al.\ (2009) reported detecting a rapid gamma-ray  flare from PKS 1510$-$089 in March 2008 using the AGILE satellite.  \par

We found significant variations in the brightness of PKS 1510$-$089 over a three month period of observations Fig.\ 10). 
Even though there is not much difference in the percentage variation in the magnitude of the source (see Table 4) in different 
bands,  the colour variations are still significant (except for V$-$R).
PKS 1510$-$089 decayed from 16.1 to 17 magnitude in the B band during our observations, which is
only $\sim$0.8 magnitude brighter than the faintest magnitude (B$_{mag}$ = 17.8) reported in the historic
LC of Liller \& Liller (1975); hence,  we observed the source in a faint phase.  \par

\noindent
{\bf BL Lac:} The object BL Lac is the archetype of its class. Observations over the  past
few decades have showed that its optical and radio emissions are highly variable
and polarized and the polarization at those widely separated  frequencies is found to be strongly correlated
(e.g., Sitko \& Schmidt 1985).  BL Lac is among the very few sources for which more than 100 years of optical
data is available in the literature (Shen 1970; Webb et al.\ 1988; Fan et al.\ 1998). An optical
variation of $\Delta$B = 5.3 mag and a possible periodicity of $\sim$14 years has been reported
for BL Lac
by Fan et al. (1998a). Very recently, Nieppola et al.\ (2009) have studied the long term variability
of the source at radio frequencies and generalized a shock model that can 
explain it. \par

We found that BL Lac was variable in all the observed passbands during our observations (Fig.\ 11). As the 
per centage variation is nearly equal in all the four passbands (although a little higher in B band), we find no significant colour variations (except perhaps B$-$V).
The average R band magnitude of the source during our
observing run is 13.95 (13.8 --14.1) which is $\sim$1.0 magnitude brighter than the faintest
magnitude reported for the source by Fan et al.\ (2000).  \par

\noindent
{\bf 3C 454.3:} This FSRQ  is among the most intense and
variable sources. The source has been detected in the flaring state in July 2007 and July 2008 at
$\gamma$-ray frequencies and those flares have been found to be well correlated with optical and
longer wavelength flares (Ghisellini et al.\ 2007; Raiteri et al.\ 2008; Villata et al.\ 2007). The
long term observational properties of 3C 454.3 at optical and radio frequencies have been well
studied through multiwavelength campaigns  (e.g., Villata et al.\ 2006, 2007). The IDV of the source 
has been recently studied by Gupta et al.\ (2008c) who reported
that the amplitude varied by $\sim$5--17 per cent during their 
observations.  \par

The source 3C 454.3 showed large flux variations in all the passbands during our observations and the per centage variation differs significantly, which leads to significant variations in the colour of source (Fig.\ 12).
The source decayed significantly from R=14.5 to 15.5, i.e.,
$\Delta$R = 1 magnitude  during our observing run of 50 days;  this is $\sim$3.5 magnitudes fainter than the
brightest magnitude (R$_{mag}$ = 12) but $\sim$1.5 magnitudes brighter than the faintest magnitude
(R$_{mag}$ = 17) reported in the source by Villata et al.\ (2006). It is very likely that we have observed
this FSRQ in a post-outburst state.  \par

\subsection{Correlated variations between colour and magnitude}
Any relationships between the variations in brightness of each of these 12 blazars and the corresponding variations
in their B$-$V, V$-$R, R$-$I and B$-$I colour indices are worth examining. Such colour-magnitude plots of the individual sources are
displayed in Figs.\ 13 -- 15. We display colour indices that are calculated by only considering
data taken by the same instrument within  a time interval of no more than 20 minutes; still,
the possibility of a rapid variation in overall flux within that interval sufficient to confound our measurements must be noted.
The individual panels show the B$-$V, V$-$R,
R$-$I and B$-$I colours plotted (in sequence from bottom to top with arbitrary offsets) with respect to V magnitude.
The straight lines shown are the best linear fit for each of  the colour indices, $CI$, against magnitude, $V$, for each
of the sources: $CI = m V + c$.  Those fitted values for the  slopes of the curves, $m$, and the
constants, $c$, are listed in Table 5.  Table 5 also gives  the linear Pearson correlation coefficients, $r$, and the corresponding
null hypothesis probability values, $p$. \par

Here a positive slope means a positive correlation between the colour index and apparent magnitude of the 
blazar, which physically means that the source tends to be bluer when it brightens or redder when it dims. A negative slope 
implies the opposite correlation between brightness and colour so that the source exhibits a redder when brighter
behaviour. We found  significant negative correlations ($p \leq 0.05$) between the V magnitudes and at least some
colour indices for the following blazars: PKS 0420$-$014 (B$-$V, V$-$R, and R$-$I); OJ 287 (B$-$V); 4C 29.45 (B$-$V);
3C 273 (B$-$V and R$-$I); PKS 1510$-$089 (B$-$V and V$-$R) and 3C 454.3 (V$-$R, R$-$I and B$-$I).
Significant positive correlations ($p \leq 0.05$) are found for: 3C 66A (V$-$R and R$-$I); S5 0716$+$714 (B$-$V,
V$-$R, R$-$I and B$-$I); 3C 273 (V$-$R); 3C 279 (V$-$R and R$-$I) and BL Lac (B$-$V and B$-$I) while the correlations
among the rest have significance less than 95$\%$. The two sources  lacking any such correlations were AO 0235$+$164 and
PKS 0735$+$178.  The only blazar in our sample to show both positive and negative correlations, depending upon the bands considered, was 3C 273.  \par

\subsection{Variability duty cycles on STV timescales}

The duty cycle (DC) for variability of a class of objects can be roughly taken to be  the fraction of them showing
significant variations on particular timescales. If $N$ denotes the total number of sources observed in a 
search for STV and $n$ denotes the number of sources that are found to be variable, then the duty cycle, as  a per centage (Gupta et al.\ 2009b), is defined as \par
\begin{eqnarray}
DC = \frac{n}{N}\times 100  
\end{eqnarray}
We detected STV in 11 out of 12 LBLs, so the DC of flux variation on STV timescales is $\sim$92 per cent.\par
We searched for variations in all of the B$-$V, V$-$R, R$-$I and B$-$I colour indices of each source. 
In our sample of a dozen blazars we found strong short-term colour variability (STCV) in most of them, using a  $>$ 0.99 probability
definition for the presence of such variations ($C > 2.57$). Based on our results here, we can categorize the
colour variations into three classes: strong STCV; partial STCV; and non-STCV, or no variation of colour with time.
The sources  that we find to  have strong STCV  are AO 0235$+$164, PKS 0420$-$014, 4C 29.45 and 3C 454.3,
as they show significant variations in all the four colour indices.
The  formally non-STCV sources during our observations were 3C 66A, OJ 287, 3C 273 and BL Lac.
The remainder of the blazars showed one to three nominally significant colour variations and 
thus could  be fairly said to have exhibited partial STCV.  If we are conservative and only consider the
four blazars with strong STCV  then the DC of colour variation  on STV timescales is $\sim$33 per cent.

\section{Discussion}

\subsection{Flux Variations}

During the period 2008 September to 2009 June we  carried out multiband photometric observations of 12 LBLs .  Comparisons of our
observations with earlier measurements of the same sources indicated that the blazars: PKS 0420$-$014, 4C 29.45 and PKS 1510$-$089 
were in relatively faint states; AO 0235$+$164, BL Lac and 3C 454.3 were possibly in post-outburst states; S5 0716$+$714 and 
3C 66A were more likely to be in pre-outburst states; while the two sources PKS 0735$+$178 and OJ 287 were in some intermediate state
during our observing run, and none of them appeared to be in outburst.  We cannot even attempt to classify  3C 273  in this fashion
because it never has shown  a large amount of optical activity:  the observed total
$\Delta$B is a modest $\sim$2 mag between 1887 and 2000 (Courvoisier et al.\ 1998; Dai et al. 2006b). We observed large flux 
variations in all the sources except 3C 273 during our observing span. 
\par
The substantial flux variations we and others have observed in these LBLs can be reasonably explained by models that involve relativistic
shocks propagating outward (e.g., Marscher \& Gear 1985; Wagner \& Witzel 1995; Marscher 1996).  The larger flares are expected
to be produced by  the emergence and motion of a new shock triggered by some strong variation in a physical quantity such as velocity, 
electron density  or magnetic field moving into and through the relativistic jet.  Smaller variations may be nicely explained
by turbulence behind a shock propagating down the jet (Marscher et al.\
1992). In a minority of blazars, gravitational micro-lensing (e.g., Gopal-Krishna \& Subramanian 1991) might be
important.  The one such case  among our sample where micro-lensing may play a role is AO 0235$+$164 which is known to have two
galaxies (at $z = 0.524$ and $z=0.851$) along our line-of-sight to it.  \par

\subsection{Colour Variations}

The variability of blazars has been extensively studied over the past two decades. Still, to the best of our knowledge,  we are reporting the most extensive search for optical short term colour 
variations (STCV), i.e., variations in colour 
with time for time periods corresponding to STV, in a sample of LBLs. 
Gu et al.\ (2006) presented monitoring of eight LBLs, seven of which are also in our sample of 12, for about six months during 2003 and 2004; their observations were only in the V, R and I bands for six of their sources, though they did also have B band coverage for two of them. \par 
Three  phenomena that could lead to colour variations with time were discussed by
Hawkins et al.\ (2002): the effect of the underlying host galaxy; colour changes in accretion disks; and colour changes from micro-lensing and we now consider each of these. The effect of the underlying host galaxy on colour variations of the
source is usually seen in the case of low luminosity AGNs such as Seyfert galaxies (M$_{B} > -22$) where variable 
seeing would include more or less of the galaxy's light along with that of the Seyfert.
Since the FSRQs are among the most luminous AGNs (M$_{B} < -23$), any
colour variation in these sources is most unlikely to arise from the underlying host galaxy.
This is  true for BL Lacs as well, in that the Doppler boosted jet emission almost invariably swamps the light from the host galaxy, thereby often making the determination of redshifts very difficult, as noted above.  \par
The STCV observed in blazar LCs might conceivably come from colour changes
arising in the accretion disc itself,  particularly for the FSRQ class, where accretion disc emission may be significant (e.g., Malkan 1983; Pian et al.\ 1999), although the BL Lacs
seem to have much weaker disc emission.
The standard model of a geometrically and optically thick
accretion disc (e.g., Shakura \& Sunyaev 1973) yields a negative temperature gradient, i.e., the disc is cooler
and redder with increasing radial distance from the central black hole.  Hence any non-linear oscillations of such a disc lead to temperature, and hence colour,
changes (Hawkins et al.\ 2002).  Microlensing of radiation from an accretion disc having the expected  radial
temperature gradient can also produce  variations in colour.  Yonehara et al.\ (1999) carried out
numerical simulations of the microlensing of accretion discs by a compact body and showed that
for an optically thick accretion disk with a radial temperature gradient colour variations are seen,
while for optically thin accretion discs such induced colour variations are unlikely to be observable.
The blazar AO 0235$+$164 showed strong STCV during our observing run and is
known to have two  foreground galaxies at $z = 0.524$ and $z = 0.851$ (Nilsson et al.\ 1996).
So at least some of the strong variations seen in this source could arise from microlensing 
 of different regions within a relativistic jet (e.g., Gopal-Krishna \& Subramanian 1991).
Since there are no microlenses known to be in front of the FSRQs  PKS 0420$-$014, 4C 29.45 and 3C 454.3, it is possible that the STCV observed in these sources arise from of colour changes 
in the accretion disc itself.  \par
The usual models for BL Lacs  attribute the vast bulk of their emission to Doppler boosted
fluxes from the relativistic jet (e.g., Scheuer \& Readhead 1979; Blandford \& K{\"o}nigl 1979).  Under these
circumstances any accretion disc radiation is always likely to be overwhelmed by that from the jet, just as the
galaxy's light also only comprises a small fraction of that observed.  Therefore jet models that can
produce different fluctuations at different colours are most likely to be able to explain the STV and colour variations 
we have detected. 
In the usual boosted synchrotron models involving shocks propagating
down the jets (e.g., Marscher \& Gear 1985; Marscher et al.\ 2008) radiation at different frequencies is
produced (or at least emerges when the jet becomes optically thin to it) at different distances behind the
shocks.  Typically higher frequency photons emerge sooner (e.g., Valtaoja et al.\  1992).
Therefore as we see a blazar brighten or fade we expect to see some differences in
the times and rates at which radiation observed at different visible colours rise and fall. \par

\subsection{Relation between colour index and flux variation}

Correlated  trends in variability between an optical colour index and brightness has been studied by quite a few
authors for a number of blazars, although usually just two colours were monitored as opposed to the four we 
usually could obtain. The historical optical observations seem to show that the correlation between
colour index and magnitude for BL Lac objects is such that their spectra become steeper
(redder) when their fluxes decrease and flatter (bluer) when they increase. In this context the blazar BL Lac is one of 
the most extensively studied objects (Carini et al. \ 1992b; Massaro et al.\ 1998; Nesci et al.\ 1998; Webb et al.\ 1998; 
Clements et al.\ 2001;  Villata et al.\ 2002, 2004; Gu et al.\ 2006).  These extensive data show that this  source usually 
follows a bluer-when-brighter trend, both in flaring and nearly steady states. A similar trend was also observed in S5 
0716$+$714 (Ghisellini et al.\ 1997; Raiteri et al.\ 2003; Gu et al.\ 2006), 3C 66A (Ghosh et al.\ 2000; Gu et al.\  2006) and OJ 287 (Carini et al.\ 1992a; Gu et al.\ 2006). Apart from the apparently general trend in BL Lac 
objects, a redder-when-brighter correlation has sometimes  been observed in the BL Lacs PKS 0735$+$178,
3C 66A and AO 0235$+$164 (Ghosh et al.\  2000).   This overall trend is easily explained if there is a shift to higher
peak frequencies during a flare (e.g., Giommi et al.\ 1999 and references therein), but there is as yet no
clear understanding as to why such a shift should occur.  The FSRQs have been much less well studied in this regard, but
a general redder-when-brighter trend seems to be present in at least two of them, 3C 454.3 and PKS 0420$-$014
(Gu et al.\ 2006 and references therein), as well as another, CTA 102 (Osterman Meyer et al.\ 2009). \par

Our data do support the general bluer-when-brighter trend for BL Lacs since three of the six BL Lacs we examined
showed this trend: 3C 66A, S5 0716$+$714 and BL Lac itself.  In our observations
AO 0235$+$164 and PKS 0735$+$178 showed no significant trend either way, and only OJ 287 showed a weak redder-when-brighter
correlation.  More importantly, our results add quite a bit of evidence in favour of the hint of a redder-when-brighter 
trend for FSRQs (Gu et al.\ 2006; Osterman Meyer et al.\ 2009). Four of the six FSRQs in our sample, PKS 0420$-$014, 
4C 29.45, PKS 1510$-$089 
and 3C 454.3 all show that trend.  The special object 3C 273 had similar negative correlations for two colour-indices but a positive
one for the other; only 3C 279 among the FSRQs showed a clear bluer-when-brighter trend, and it was one of the two objects 
for which we lacked B band data. \par
This possibility  that the colour index variations with brightness typically follow different trends in FSRQs and BL Lacs 
is intriguing.  If a general result, it would indicate that the physical conditions are different in the two types of 
blazars in some important way.  This perhaps would not be too surprising, in that most BL Lacs are believed to be beamed, 
but intrinsically weak, FR I radio  galaxies (e.g., Padovani \& Urry 1992), while the parent population of the FSRQs are 
the more powerful FR II radio galaxies (e.g., Maraschi \& Rovetti 1994).   On the other hand, the hypothesis that FR II 
and FR I radio galaxies  have fundamentally similar engines and that the difference between them arises merely from 
difference in their jet powers and the gases through which they flow on the kpc-scale (e.g., 
Bicknell 1994; Maraschi \& Rovetti 1994)  is well supported, particularly  by the existence of HYMORS -- HYbrid MOrphology
Radio Sources, in which one side of the source has an FR I morphology, while the other shows a clear FR II structure (e.g., Gopal-Krishna \& Wiita 2000).  Still, there do seem to be some differences in the nuclear UV properties of the two classes (e.g., 
Chiaberge et al.\ 2002) so this dichotomy is not completely clear.\par  
So it is very important to investigate whether the claim that the BL Lacs and FSRQs follow different trends 
for colour index against brightness is correct.  While the preponderance of data, both in 
BL Lacs and FSRQs, is with these opposite trends the numbers are small and exceptions are present in both groups.
All the optical observations of the source BL Lac shows that it follows a bluer-when-brighter trend and our
observations also revealed the same trend. A similar trend has also been commonly observed in S5 0716$+$714 and 
OJ 287. Our literature search shows the general steepening trend with increase in brightness of FSRQs in
3C 454.3 and PKS 0420$-$014  which we have also found in our analysis results.  In PKS 0735$+$178 and 3C 66A, Ghosh et al.\
(2000) report the steepening trend with increases in brightness while Gu et al.\ (2006) report the opposite trend.
During our observations, both of them show a bluing trend with increase in flux although the correlation
is weak in the case of PKS 0735$+$178. We stress that the amount of data for FSRQs is quite limited and additional 
observations could easily erase the nominal trend we find for redder-when-brighter behaviour 
among them.  \par
We suggest that both of these behaviours can be accommodated within shock-in-jet models.
When an outgoing shock strikes a region with larger electron populations, stronger magnetic fields
or a turbulent cell within the underlying flow, the emissivity from that region will rise. The radiation emitted at 
higher frequencies will typically emerge sooner while the maximum flux at somewhat lower frequencies will be
seen later (e.g., Valtaoja et al.\ 2002).  Therefore during the early phase of a rise in flux one is more likely 
to see a bluer colour. However, an observation later during the same flare might show a more enhanced redder band as the
bluer one might have stopped rising as fast or may even have passed its peak. The former situation should be
more likely to be seen than the latter, yielding a predominance of bluer-when-brighter situations,  particularly when the jet is completely dominating the flux, as is the case for BL Lacs.  Whereas for FSRQs there can be a significant disc component to the optical emission and this contribution is stronger toward the UV where the disc temperature peaks.  Then an increase in synchrotron emission from the jet, which usually peaks in the IR  for LBLs, would naturally yield  a redder-when-brighter trend (e.g. Gu et al.\ 2006).  Our one counterexample among the FSRQs, 3C 279, can sometimes have quite weak lines; since we seem to have observed it in a low-state, it may well have been in that
circumstance.  Then it could easily have been disc dominated, and thus shown a bluer-when-brighter trend, though it might behave in the opposite manner in a phase when its jet component is stronger (e.g., Pian et al.\ 1999).\par

\section{Conclusions}

We have carried out multiband optical photometry of a sample of 12 LBLs during 2008 September through 2009 June. The
results of our study can be summarized as follows: \\
1) All sources except 3C 273 showed strong flux variations on timescales of months, as expected.
The amplitude of
variation was $\sim$0.3 mag in PKS 0420$-$014, BL Lac and PKS 0735$+$178, while for AO 0235$+$164 and
S5 0716$+$714 the variability amplitude we saw exceeded 2 magnitudes.  For the remaining  eight blazars
the amplitude of variability exceeded 0.9 magnitudes. \\
2) We found strong  STCV in at least one colour index for 8 of the 12 sources, with 3C 66A,
OJ 287, 3C 273 and BL Lac as the exceptions, but only 4 showed it in all 4 bands. \\
3) The duty cycles for flux and colour variations on STV timescales in LBLs are $\sim$92 per cent  and $\sim$33 per cent, respectively. \\
4) Three out of six BL Lacs followed the bluer-when-brighter trend that had  been rather well established
in earlier studies; AO 0235+164 was achromatic; however, both OJ 287 and PKS 0735$+$178 were
redder-when-brighter.
5) Four of the six FSRQs showed the opposite, redder-when-brighter, trend, which had been suggested earlier.
Two colour indices went in this direction for 3C 273 but one went the other way; the only clear violator of this  trend
was 3C 279. Our new data substantially increases the evidence in favour of this proposition, but by no means
establish it. Substantial additional multi-colour observations of this class of blazars would be needed to do
so and they are on our agenda.

\section*{Acknowledgments} We thank the anonymous referee for numerous suggestions that
led to an improved presentation.   BR is thankful to Mr.\ Tarun Chhikara 
for  help while finalizing the text. PJW is grateful for hospitality at ARIES.
This research was partially supported by Scientific Research Fund of the
Bulgarian Ministry of Education and Sciences (BIn - 13/09 and DO02-340/08) and by 
Indo $-$ Bulgaria bilateral scientific exchange project INT/Bulgaria/B$-$5/08 funded 
by DST, India.

\begin{table*}
\caption{ Details of Telescopes and Instruments Used}
\textwidth=6.0in
\textheight=9.0in
\vspace*{0.2in}
\noindent
\begin{tabular}{llllll} \hline
Site:        &ARIES Nainital & NAO Rozhen         & NAO Rozhen            & NAO Rozhen         & AO Belogradchik             \\\hline
Telescope:   &1.04-m RC Cassegrain           & 2-m Ritchey-Chr\'etien  & 50/70-cm Schmidt & 60-cm Cassegrain   & 60-cm Cassegrain               \\
CCD model:   & Wright 2K CCD                 & PI VersArray:1300B      & SBIG STL-11000M  & FLI PL09000        & FLI PL09000                    \\
Chip size:   & $2048\times2048$ pixels                               & $1340\times1300$ pixels & $4008\times2672$ pixels & $3056\times3056$ pixels & $3056\times3056$ pixels   \\
Pixel size:  &$24\times24$ $\mu$m             & $20\times20$ $\mu$m     & $9\times9$ $\mu$m & $12\times12$ $\mu$m    & $12\times12$ $\mu$m            \\
Scale:       &0.37\arcsec/pixel               &0.258\arcsec/pixel       & 1.079\arcsec/pixel& 0.330\arcsec/pixel$^{\rm a}$   & 0.330\arcsec/pixel$^{\rm a}$   \\
Field:       & $13\arcmin\times13\arcmin$     &$5.76\arcmin\times5.59\arcmin$ & $72.10\arcmin \times 48.06 \arcmin$ & $16.8\arcmin\times16.8\arcmin$ & $16.8\arcmin\times16.8\arcmin$ \\
Gain:        &10 $e^-$/ADU                    &1.0 $e^-$/ADU             & 0.84 $e^-$/ADU    & 1.0 $e^-$/ADU    & 1.0 $e^-$/ADU                  \\
Read Out Noise:         &5.3 $e^-$ rms                   &2.0 $e^-$ rms             & 13.0 $e^-$ rms    & 8.5 $e^-$ rms    & 8.5 $e^-$ rms                  \\
Binning used:&$2\times2$                      &$1\times1$                & $1\times1$        & $2\times2$       & $3\times3$                     \\
Typical seeing : & 1\arcsec to 2.8\arcsec  & 1.5\arcsec to 3.5\arcsec & 1.5\arcsec to 3.5\arcsec & 1.5\arcsec to 3.5\arcsec & 1.5\arcsec to 3.5\arcsec   \\\hline
\end{tabular} \\
\noindent
$^{\rm a}$ With a binning factor of $1\times1$
\end{table*}

\begin{table*}
\caption{Observation log of optical photometric observations}
\textwidth=6.0in
\textheight=9.0in
 \centering
\vspace*{0.2in}
\noindent
\begin{tabular}{lccccll} \hline
Source        &$\alpha_{2000.0}$&$\delta_{2000.0}$& Date of Observation & Telescope  & Filters  & Data Points \\
(z)          &(hh mm ss)      & (hh mm ss)        &(yyyy mm dd)         &            &          &             \\\hline
3C 66A        & 02 22 39.61    & +43 02 07.80     & 2008 10 20          &   A  &B,V,R,I    &1,1,1,1  \\
(0.444)       &                &                  & 2008 10 22          &   A  &B,V,R,I    &1,1,1,1  \\
              &                &                  & 2008 10 23          &   A  &B,V,R,I    &1,1,1,1  \\
              &                &                  & 2008 10 24          &   A  &B,V,R,I    &1,1,1,1  \\
              &                &                  & 2008 10 25          &   A  &B,V,R,I    &1,1,1,1  \\
              &                &                  & 2008 10 26          &   A  &B,V,R,I    &1,1,1,1  \\
              &                &                  & 2008 10 27          &   A  &B,V,R,I    &1,1,1,1  \\
              &                &                  & 2008 10 28          &   A  &B,V,R,I    &1,1,1,1  \\
              &                &                  & 2008 10 30          &   A  &B,V,R,I    &1,1,1,1  \\
              &                &                  & 2009 01 20          &   A  &B,V,R,I    &1,1,1,1  \\
              &                &                  & 2009 01 22          &   A  &B,V,R,I    &1,1,1,1  \\
              &                &                  & 2009 02 02          &   A  &B,V,R,I    &1,1,1,1  \\
AO 0235+164   & 02 38 39.93    & +16 36 59.27     & 2008 09 04          &   A  &B,V,R,I    &1,1,1,1  \\
(0.94)        &                &                  & 2008 10 20          &   A  &B,V,R,I    &1,1,1,1  \\
              &                &                  & 2008 10 22          &   A  &B,V,R,I    &1,1,1,1  \\
              &                &                  & 2008 10 23          &   A    &B,V,R,I    &1,1,1,1  \\
              &                &                  & 2008 10 24          &   A    &B,V,R,I    &1,1,1,1  \\
              &                &                  & 2008 10 25          &   A    &B,V,R,I    &1,1,1,1  \\
              &                &                  & 2008 10 26          &   A    &B,V,R,I    &1,1,1,1  \\
              &                &                  & 2008 10 27          &   A    &B,V,R,I    &1,1,1,1  \\
              &                &                  & 2008 10 28          &   A    &B,V,R,I    &1,1,1,1  \\
              &                &                  & 2008 10 30          &   A    &B,V,I      &1,1,1    \\
              &                &                  & 2009 01 20          &   A    &B,V,R,I    &1,1,1,1  \\
              &                &                  & 2009 01 22          &   A    &B,V,R,I    &1,1,1,1  \\
PKS 0420$-$014  & 04 23 15.73    & -01 20 32.70     & 2008 10 23        &   A     &B,V,R,I   & 1,1,1,1  \\
(0.915)       &                &                  & 2008 10 24          &   A     &B,V,R,I   & 1,1,1,1  \\
              &                &                  & 2008 10 26          &   A     &B,V,R,I   & 1,1,1,1  \\
              &                &                  & 2009 01 20          &   A     &B,V,R,I   & 1,1,1,1  \\
              &                &                  & 2009 01 22          &   A     &B,V,R,I   & 1,1,1,1  \\
              &                &                  & 2009 02 02          &   A     &B,V,R,I   & 1,1,1,1  \\
S5 0716+714   & 07 21 53.45    & +71 20 36.35     & 2008 10 24          &   A     &B,V,R,I   & 1,1,1,1  \\
(0.31$\pm$0.08)&                &                  & 2008 10 26          &   A     &B,V,R,I   & 1,1,1,1  \\
              &                &                  & 2009 02 25          &   A     &B,V,R,I   & 1,1,1,1  \\
              &                &                  & 2009 03 26          &   C     &B,V,R,I   & 2,2,2,2  \\
              &                &                  & 2009 04 16          &   C     &B,V,R,I   & 2,2,2,2  \\
              &                &                  & 2009 04 17          &   B     &U,B,V,R,I & 2,2,2,2,2 \\
              &                &                  & 2009 05 19          &   C     &B,V,R,I   & 2,2,2,2  \\
              &                &                  & 2009 05 20          &   C     &B,V,R,I   & 2,2,2,2  \\
              &                &                  & 2009 05 21          &   C     &B,V,R,I   & 2,2,2,2  \\
PKS 0735+178  & 07 38 07.39    & +17 42 18.99     & 2009 01 20          &   A    &B,V,R,I   & 1,1,1,1  \\
(0.424)       &                &                  & 2009 01 22          &   A    &B,V,R,I   & 1,1,1,1  \\
              &                &                  & 2009 03 25          &   D    &V,R,I     & 2,2,2    \\
              &                &                  & 2009 03 26          &   D    &V,R       & 1,1      \\
              &                &                  & 2009 04 16          &   D    & R,I      & 2,1       \\
OJ 287        & 08 54 48.87    & +20 06 30.64     & 2008 10 24          &   A    &B,V,R,I   & 1,1,1,1   \\
(0.306)       &                &                  & 2008 10 26          &   A    &B,V,R,I   & 1,1,1,1   \\
              &                &                  & 2009 01 20          &   A    &B,V,R,I   & 1,1,1,1   \\
              &                &                  & 2009 01 22          &   A    &B,V,R,I   & 1,1,1,1   \\
              &                &                  & 2009 03 24          &  C     &B,V,R,I   &1,2,2,2    \\
              &                &                  & 2009 03 25          &  C     &B,V,R,I   &1,2,2,2    \\
              &                &                  & 2009 03 26          &  D,   C      &B,V,R,I&3,4,4,4 \\
              &                &                  & 2009 04 01          &   A    & B,V,R,I   & 1,1,1,1   \\
              &                &                  & 2009 04 15          &  D,   C      &B,V,R,I&2,2,2,2  \\
              &                &                  & 2009 04 16          &  C      &B,V,R,I   & 4,4,4,4   \\
              &                &                  & 2009 04 17          &   A     &B,V,R,I   & 1,1,1,1  \\
              &                &                  & 2009 04 18          &   A     &B,V,R,I   & 1,1,1,1  \\
              &                &                  & 2009 04 19          &   A     &B,V,R,I   & 1,1,1,1  \\
              &                &                  & 2009 05 17          &   C     &B,V,R,I   & 2,2,2,7  \\
              &                &                  & 2009 05 20          &   C     &V,R,I     & 1,1,1  \\
              &                &                  & 2009 05 21          &   C     &B,V,R,I   & 2,2,2,2  \\
              &                &                  & 2009 05 25          &   A     &B,V,R,I   & 1,1,1,1  \\
              &                &                  & 2009 05 26          &   A     &B,V,R,I   & 1,1,1,1  \\\hline
\end{tabular}
\end{table*}

\begin{table*}
{ Table 2. continued ...}
\textwidth=6.0in
\textheight=9.0in

\vspace*{0.2in}
\noindent
\begin{tabular}{lccccll} \\\hline
4C 29.45      & 11 59 32.07    & +29 14 42.00     & 2009 03 25          &   D    &V,R,I      &2,2,2    \\
(0.729)       &                &                  & 2009 03 26          &   D    &B,V,R,I    &2,2,2,2  \\
              &                &                  & 2009 04 15          &   D    &V,R,I      &2,3,4    \\
              &                &                  & 2009 04 16          &   D    &V,R,I      &2,2,2    \\
              &                &                  & 2009 04 17          &   A    &B,V,R,I    &1,1,1,1  \\
              &                &                  & 2009 04 18          &   A    &B,V,R,I    &1,1,1,1  \\
              &                &                  & 2009 04 23          &   B    &B,V,R,I    &2,2,2,2  \\
              &                &                  & 2009 04 24          &   B    &B,V,R,I    &2,2,2,2  \\
              &                &                  & 2009 05 15          &  C     &B,V,R,I    &2,2,2,2  \\
              &                &                  & 2009 05 17          &  C     &V,R,I      &1,1,1  \\
              &                &                  & 2009 05 21          &  C,   D&V,R,I   &2,2,2  \\
              &                &                  & 2009 05 22          &  C     &V,R,I      &1,1,1  \\
              &                &                  & 2009 05 23          &  C,   A,   D &B,V,R,I  &1,3,3,3  \\
              &                &                  & 2009 05 25          &  C,   A      & B,V,R,I   &1,2,2,2   \\
              &                &                  & 2009 05 26          &  C     &V,R,I      &1,1,1  \\
              &                &                  & 2009 05 27          &   A    &B,V,R,I    &1,1,1,1  \\
              &                &                  & 2009 06 14          &   E    &V,R,I      &2,2,2    \\
3C 273        & 12 29 06.69    & +02 03 08.58     & 2009 20 01          &   A    & B,V,R,I   & 1,1,1,1   \\
(0.1575)      &                &                  & 2009 22 01          &   A    & B,V,R,I   & 1,1,1,1   \\
              &                &                  & 2009 25 02          &   A    & B,V,R,I   & 1,1,1,1   \\
              &                &                  & 2009 24 03          &   A    & B,V,R,I   & 1,1,1,1   \\
              &                &                  & 2009 25 03          &   D    &U,B,V,R,I  & 2,2,2,2,2 \\
              &                &                  & 2009 26 03          &   D    &U,B,V,R,I  & 2,2,2,2,2 \\
              &                &                  & 2009 04 01          &   A    & B,V,R,I   & 1,1,1,1   \\
              &                &                  & 2009 04 15          &   D    & U,B,V,R,I  & 2,2,2,2,2 \\
              &                &                  & 2009 04 16          &   D    & U,B,V,R,I  & 2,2,2,2,2 \\
              &                &                  & 2009 04 17          &   B    & U,B,V,R,I  & 2,2,2,2,2 \\
              &                &                  & 2009 04 18          &   A    & B,V,R,I   & 1,1,1,1   \\
              &                &                  & 2009 04 19          &   A    & B,V,R,I   & 1,1,1,1   \\
              &                &                  & 2009 04 27          &   A    & B,V,R,I   & 1,1,1,1   \\
              &                &                  & 2009 05 15          &   D    & B,V,R,I   & 3,2,2,2   \\
              &                &                  & 2009 05 21          &   D,   C       & B,V,R,I   & 3,3,3,3   \\
              &                &                  & 2009 05 23          &   A    & B,V,R,I   & 1,1,1,1   \\
              &                &                  & 2009 05 25          &   A,   D    & B,V,R,I   & 3,3,3,3   \\
              &                &                  & 2009 05 26          &   A,   D    & U,B,V,R,I & 1,2,3,3,3   \\
              &                &                  & 2009 05 27          &   A    & B,V,R,I   & 1,1,1,1   \\
              &                &                  & 2009 06 03          &   E    & B,V,R,I  & 2,2,2,2   \\
              &                &                  & 2009 06 14          &   E    & B,V,R,I  & 6,4,4,4   \\
              &                &                  & 2009 06 15          &   E    & B,V,R,I  & 6,6,4,6   \\
              &                &                  & 2009 06 16          &   E    & B,V,R,I  & 8,4,6,6   \\
3C 279        & 12 56 11.17    & $-$05 47 21.52     & 2009 03 25        &   D    &V,R,I      &2,1,2      \\
(0.5362)      &                &                  & 2009 03 26          &   D    &V,R,I      &2,3,2   \\
              &                &                  & 2009 04 15          &   D    &V,R,I      &2,2,2   \\
              &                &                  & 2009 04 16          &   D    &V,R,I      &2,2,2   \\
              &                &                  & 2009 05 15          &   D    &V,R,I      &1,1,1  \\
              &                &                  & 2009 05 19          &   D    & V,R,I     &1,1,1  \\
              &                &                  & 2009 05 21          &  C     & V,R,I     &1,2,2  \\
              &                &                  & 2009 05 25          &   D    & V,R,I     &1,1,1  \\
              &                &                  & 2009 05 26          &   D    & V,R,I     &1,1,1  \\
              &                &                  & 2009 06 14          &   E    & V,R,I     &4,4,4 \\
              &                &                  & 2009 06 15          &   E    & V,R       & 4,2    \\
PKS 1510$-$089  & 15 12 50.53    & -09 05 58.99     & 2009 04 17        &   A    &B,V,R,I   & 1,1,1,1  \\
(0.36)        &                &                  & 2009 04 19          &   A    &B,V,R,I   & 1,1,1,2  \\
              &                &                  & 2009 04 27          &   A    &B,V,R,I   & 1,1,1,1  \\
              &                &                  & 2009 05 25          &   A    &B,V,R,I   & 1,1,1,1  \\
              &                &                  & 2009 05 27          &   A    &B,V,R,I   & 1,1,1,1  \\
              &                &                  & 2009 06 14          &   A    &B,V,R,I   & 1,1,1,1  \\
              &                &                  & 2009 06 21          &   A    &  V,R,I   &   1,1,1  \\
              &                &                  & 2009 06 24          &   A    &B,V,R,I   & 1,1,1,1  \\\hline
\end{tabular}
\end{table*}

\begin{table*}
{ Table 2. continued ...}
\textwidth=6.0in
\textheight=9.0in

\vspace*{0.2in}
\noindent
\begin{tabular}{lccccll} \\\hline
BL Lac        & 22 02 42.29    & +42 16 39.98     & 2008 09 04          &   A   & B,V,R,I   & 1,1,1,1   \\
(0.069)       &                &                  & 2008 09 07          &   A    & B,V,R,I   & 1,1,1,1   \\
              &                &                  & 2008 09 08          &   A    & R         & 1         \\
              &                &                  & 2008 09 10          &   A    & B,V,R,I   & 1,1,1,1   \\
              &                &                  & 2008 10 23          &   A    & B,V,R,I   & 1,1,1,1   \\
              &                &                  & 2008 10 24          &   A    & B,V,R,I   & 1,1,1,1   \\
              &                &                  & 2008 10 25          &   A    & B,V,R,I   & 1,1,1,1   \\
              &                &                  & 2008 10 26          &   A    & B,V,R,I   & 1,1,1,1   \\
              &                &                  & 2008 10 27          &   A    & B,V,R,I   & 1,1,1,1   \\
              &                &                  & 2008 10 30          &   A    & B,V,R,I   & 1,1,1,1   \\
              &                &                  & 2009 04 17          &   B    &U,B,V,R,I  &2,2,2,5,2  \\
3C 454.3      & 22 53 57.75    & +16 08 53.56     & 2008 09 06          &   A    &R          &1        \\
(0.8590)      &                &                  & 2008 09 07          &   A    &B,V,R,I    &1,1,1,1   \\
              &                &                  & 2008 09 09          &   A    &B,V,R,I    &1,1,1,1  \\
              &                &                  & 2008 10 20          &   A    &B,V,R,I    &1,1,1,1  \\
              &                &                  & 2008 10 22          &   A    &B,V,R,I    &1,1,1,1  \\
              &                &                  & 2008 10 23          &   A    &B,V,R,I    &1,1,1,1  \\
              &                &                  & 2008 10 24          &   A    &B,V,R,I    &1,1,1,1  \\
              &                &                  & 2008 10 25          &   A    &B,V,R,I    &1,1,1,1  \\
              &                &                  & 2008 10 26          &   A    &B,V,R,I    &1,1,1,1  \\
              &                &                  & 2008 10 27          &   A    &B,V,R,I    &1,1,1,1  \\
              &                &                  & 2008 10 28          &   A    &B,V,R,I    &1,1,1,1  \\
              &                &                  & 2008 10 29          &   A    &B,V,R,I    &1,1,1,1  \\
              &                &                  & 2008 10 30          &   A    &B,V,R,I    &1,1,1,1  \\\hline
\end{tabular} \\
A  : 1.04 meter Sampuranand Telescope, ARIES, Nainital, India  \\
B  : 2-m Ritchey-Chretien Telescope at National Astronomical Observatory Rozhen, Bulgaria \\
C  : 50/70-cm Schmidt Telescope at National Astronomical Observatory, Rozhen, Bulgaria   \\
D  : 60-cm Cassegrain Telescope at Astronomical Observatory Belogradchik, Bulgaria \\
E   : 60-cm Cassegrain Telescope at National Astronomical Observatory Rozhen, Bulgaria \\
\end{table*}

\begin{table*}
\caption{ Standard stars in the blazar fields }

\begin{tabular}{lcccccl} \hline
Source      & Standard   &  B magnitude  &  V magnitude   &  R magnitude   &  I magnitude   & References$^a$    \\
Name        & star       &  (error)      &   (error)      &  (error)       &  (error)       &              \\\hline
3C 66A      & 1          & 14.13(0.02)  & 13.65(0.02)   & 13.36(0.01)   & 13.05(0.02)   & 5            \\
            & 2          & 15.82(0.03)  & 14.82(0.03)   & 14.28(0.04)   & 13.75(0.03)   & 5            \\
            & 3          & 16.36(0.13)  & 15.89(0.04)   & 15.46(0.12)   & 15.05(0.13)   & 5            \\
            & 4          &              & 12.79(0.04)   & 12.70(0.04)   & 12.59(0.04)   & 6            \\
            & 5          &              & 14.18(0.05)   & 13.62(0.05)   & 13.10(0.05)   & 6            \\
            & 6          & 16.77(0.08)  & 16.10(0.08)   &               &               & 7            \\
AO 0235+164 & 1          & 13.59(0.04)  & 13.03(0.03)   & 12.69(0.02)   & 12.35(0.03)   & 1           \\
            & 2          & 13.55(0.02)  & 12.71(0.02)   & 12.23(0.02)   & 11.79(0.02)   & 1           \\
            & 3          & 13.68(0.02)  & 12.92(0.02)   & 12.48(0.03)   & 12.08(0.03)   & 1           \\
            & 6          &              & 14.02(0.05)   & 13.64(0.04)   & 13.30(0.07)   & 6           \\
            & 8          & 18.22(0.14)  & 16.58(0.12)   & 15.79(0.10)   & 14.94(0.15)   & 1            \\
            & C1         &              & 14.78(0.05)   & 14.23(0.05)   & 13.76(0.08)   & 6            \\
PKS 0420$-$014& 1          & 13.02(0.03)  & 12.45(0.02)   & 12.09(0.03)   &               & 4           \\
            & 2          & 13.60(0.01)  & 13.10(0.02)   & 12.80(0.02)   & 12.44(0.03)   & 5           \\
            & 3          & 13.97(0.03)  & 13.28(0.01)   & 12.89(0.01)   & 12.50(0.04)   & 5           \\
            & 4          & 15.72(0.02)  & 14.91(0.03)   & 14.47(0.01)   & 14.01(0.05)   & 5           \\
            & 5          &              & 14.96(0.03)   & 14.37(0.03)   &               & 4           \\
            & 6          & 16.03(0.03)  & 15.18(0.03)   & 14.70(0.03)   &               & 4           \\
            & 7          &              & 15.31(0.03)   & 14.91(0.03)   &               & 4           \\
            & 8          &              & 15.99(0.03)   & 15.46(0.03)   &               & 4           \\
            & 9          &              & 16.29(0.03)   & 15.58(0.04)   &               & 4           \\
S5 0716+714 & 1          & 11.54(0.01)  & 10.99(0.02)   & 10.63(0.01)   &               & 2            \\
            & 2          & 12.02(0.01)  & 11.46(0.01)   & 11.12(0.01)   & 10.92(0.04)   & 2, 3         \\
            & 3          & 13.04(0.01)  & 12.43(0.02)   & 12.06(0.01)   & 11.79(0.05)   & 2, 3         \\
            & 4          & 13.66(0.01)  & 13.19(0.02)   & 12.89(0.01)   &               & 2            \\
            & 5          & 14.15(0.01)  & 13.55(0.02)   & 13.18(0.01)   & 12.85(0.05)   & 2, 3         \\
            & 6          & 14.24(0.01)  & 13.63(0.02)   & 13.26(0.01)   & 12.97(0.04)   & 2, 3         \\
            & 7          & 14.55(0.01)  & 13.74(0.02)   & 13.32(0.01)   &               & 2            \\
            & 8          & 14.70(0.01)  & 14.10(0.02)   & 13.79(0.02)   &               & 2            \\
PKS 0735+178& A          & 13.87(0.04)  & 13.40(0.04)   & 13.14(0.05)   & 12.85(0.06)   & 1            \\
            & C          & 15.48(0.05)  & 14.40(0.05)   & 13.87(0.06)   & 13.33(0.07)   & 1            \\
            & D          & 16.48(0.10)  & 15.80(0.07)   & 15.45(0.06)   & 15.16(0.08)   & 1            \\
OJ 287      & 2          & 13.45(0.04)  & 12.80(0.04)   & 12.46(0.05)   & 12.06(0.07)   & 1            \\
            & 4          & 15.01(0.06)  & 14.14(0.05)   & 13.72(0.06)   & 13.23(0.07)   & 1             \\
            & 10         & 15.01(0.05)  & 14.56(0.04)   & 14.26(0.06)   & 13.94(0.09)   & 1            \\
            & 11         & 15.47(0.07)  & 14.96(0.05)   & 14.67(0.07)   & 14.29(0.09)   & 1            \\
4C 29.45    & 1          & 14.01(0.06)  & 13.39(0.05)   & 13.01(0.02)   &               & 4            \\
            & 13         & 16.02(0.04)  & 15.36(0.04)   & 14.97(0.04)   & 14.62(0.07)   & 1            \\
            & 14         & 16.41(0.05)  & 15.89(0.09)   & 15.53(0.08)   & 15.16(0.18)   & 1            \\
            & 15         & 17.14(0.05)  & 16.60(0.05)   & 16.30(0.04)   & 15.88(0.30)   & 1            \\
3C 273      & C          & 12.85(0.05)  & 11.87(0.04)   & 11.30(0.04)   & 10.74(0.04)   & 1            \\
            & D          & 13.17(0.05)  & 12.68(0.04)   & 12.31(0.04)   & 11.99(0.06)   & 1            \\
            & E          & 13.33(0.07)  & 12.69(0.04)   & 12.27(0.05)   & 11.84(0.04)   & 1            \\
            & G          & 14.12(0.05)  & 13.56(0.05)   & 13.16(0.05)   & 12.83(0.05)   & 1            \\
3C 279      & 1          & 16.81(0.05)  & 15.94(0.03)   & 15.45(0.03)   & 14.99(0.04)   & 8            \\
            & 2          & 13.02(0.02)  & 12.42(0.02)   & 12.05(0.02)   & 11.69(0.02)   & 8            \\
            & 3          & 15.45(0.07)  & 14.90(0.03)   & 14.53(0.05)   & 14.18(0.09)   & 8            \\
            & 4          & 13.75(0.01)  & 13.00(0.01)   & 12.57(0.03)   & 12.17(0.01)   & 8            \\
            & 6          & 13.43(0.01)  & 12.81(0.01)   & 12.45(0.03)   & 12.12(0.01)   & 8            \\
            & 7          & 16.53(0.05)  & 15.66(0.03)   & 15.13(0.02)   &               & 4            \\
PKS 1510$-$089& 1          & 12.23(0.02)  & 11.60(0.03)   & 11.23(0.03)   & 10.87(0.02)   & 5           \\
            & 2          & 13.74(0.02)  & 13.26(0.03)   & 12.95(0.03)   & 12.60(0.06)   & 5           \\
            & 3          & 15.16(0.04)  & 14.44(0.06)   & 13.98(0.09)   & 13.54(0.05)   & 5           \\
            & 4          & 15.36(0.06)  & 14.72(0.06)   & 14.34(0.05)   & 13.87(0.05)   & 5           \\
            & 5          & 15.43(0.05)  & 14.70(0.05)   & 14.35(0.05)   &               & 4           \\
            & 6          & 16.09(0.04)  & 15.16(0.02)   & 14.61(0.02)   &               & 4           \\\hline
\end{tabular}
\end{table*}
\begin{table*}
{Table 3. continued ... }

\begin{tabular}{lcccccl} \hline
BL Lac      & B          & 14.52(0.04)  & 12.78(0.04)   & 11.93(0.05)   & 11.09(0.06)   & 1            \\
            & C          & 15.09(0.03)  & 14.19(0.03)   & 13.69(0.03)   & 13.23(0.04)   & 1            \\
            & H          & 15.68(0.03)  & 14.31(0.05)   & 13.60(0.03)   & 12.93(0.04)   & 1            \\
            & K          & 16.26(0.05)  & 15.44(0.03)   & 14.88(0.05)   & 14.34(0.10)   & 1            \\
3C 454.3    & A          & 16.85(0.05)  & 15.86(0.09)   & 15.32(0.09)   & 14.80(0.06)   & 6, 9, 10        \\
            & B          & 15.87(0.02)  & 15.21(0.06)   & 14.73(0.05)   & 14.31(0.05)   & 6, 9, 10       \\
            & C          & 15.18(0.02)  & 14.43(0.02)   & 13.98(0.02)   & 13.51(0.02)   & 9, 10, 11       \\
            & D          & 14.94(0.02)  & 13.85(0.02)   & 13.22(0.01)   & 12.63(0.01)   & 9, 10, 11       \\
            & E          & 17.10(0.14)  & 15.76(0.09)   & 14.92(0.08)   & 14.26(0.08)   & 6, 9, 10        \\
            & F          & 16.06(0.11)  & 15.21(0.11)   & 14.83(0.03)   &               & 4, 9, 10        \\
            & G          & 16.28(0.08)  & 15.42(0.08)   & 14.83(0.02)   &               & 4, 9, 10         \\
            & H          & 14.62(0.02)  & 13.65(0.04)   & 13.10(0.04)   & 12.58(0.04)   & 6, 9, 10         \\
            & C1         &              & 15.67(0.06)   & 15.27(0.06)   & 14.71(0.06)   & 6            \\\hline

\end{tabular} \\
$^a$1. Smith et al. 1985; 2. Villata et al. 1998; 3. Ghisellini et al. 1997; 4. Raiteri et al. 1998; 5. Smith et al. 1998;
6. Fiorucci et al. 1996; 7. Craine et al. 1975; 8. http://quasar.colgate.edu/$\sim$tbalonek/optical/3C279compstars.gif~;
9. Craine 1977; 10. Angione R.J., 1971, AJ 76, 412;
11. http://www.lsw.uni-heidelberg.de/projects/extragalactic/charts/2251+158.html
\end{table*}

\begin{table*}
\caption{ Results of short-term variability observations  }

\begin{tabular}{lclrcccrrl} \hline
Source Name & Duration of Observation  & Band &  N & $\sigma$(BL $-$ S$_{A}$) & $\sigma$(BL - S$_{B}$) & $\sigma$(S$_{A}$ - S$_{B}$) & C  & A & Variable\\
            & (yyyy mm dd) to (yyyy mm dd) &  &                        &                      &                     &          &          &  \\\hline
3C 66A      & 2008 10 20 - 2009 02 02  & B    & 12 & 0.286        & 0.279            & 0.015          & 19.05    & 77.2     & V       \\
            &                          & V    & 12 & 0.292        & 0.283            & 0.014          & 19.91    & 81.8     & V       \\
            &                          & R    & 12 & 0.284        & 0.272            & 0.016          & 17.59    & 78.0     & V       \\
            &                          & I    & 12 & 0.274        & 0.281            & 0.057          & 4.89     & 75.5     & V       \\
            &                          & B$-$V$^*$ & 12 &              &                  &                & 1.46     & 6.1      & NV      \\
            &                          & V$-$R  & 12 &              &                  &                & 0.96     & 4.9      & NV      \\
            &                          & R$-$I  & 12 &              &                  &                & 1.14     & 4.2      & NV      \\
            &                          & B$-$I  & 12 &              &                  &                & 1.37     & 7.4      & NV      \\
AO 0235+164 & 2008 09 04 - 2009 01 22  & B    & 12 & 0.744        & 0.739            & 0.012          & 60.00    & 226.4    & V        \\
            &                          & V    & 12 & 0.736        & 0.740            & 0.008          & 92.63    & 226.3    & V       \\
            &                          & R    & 11 & 0.758        & 0.758            & 0.006          & 133.64   & 222.7    & V       \\
            &                          & I    & 12 & 0.697        & 0.697            & 0.007          & 102.98   & 224.1    & V       \\
            &                          & B$-$V  & 12 &              &                  &                & 6.49     & 20.8     & V       \\
            &                          & V$-$R  & 11 &              &                  &                & 4.96     & 12.2     & V       \\
            &                          & R$-$I  & 11 &              &                  &                & 24.21    & 55.1     & V       \\
            &                          & B$-$I  & 12 &              &                  &                & 12.94    & 68.0     & V       \\
PKS 0420$-$014& 2008 10 23 - 2009 02 02  & B    & 6  & 0.119        & 0.100            & 0.034          & 3.26     & 23.5     & V       \\
            &                          & V    & 6  & 0.188        & 0.173            & 0.022          & 8.15     & 43.9     & V       \\
            &                          & R    & 6  & 0.271        & 0.250            & 0.025          & 10.49    & 64.9     & V       \\
            &                          & I    & 6  & 0.293        & 0.287            & 0.009          & 32.73    & 77.1     & V       \\
            &                          & B$-$V  & 6  &              &                  &                & 4.83     & 39.4     & V       \\
            &                          & V$-$R  & 6  &              &                  &                & 4.59     & 24.2     & V       \\
            &                          & R$-$I  & 6  &              &                  &                & 4.03     & 23.6     & V       \\
            &                          & B$-$I  & 6  &              &                  &                & 7.30     & 72.8     & V       \\
S5 0716+714 & 2008 10 24 - 2009 04 17  & B    & 9  & 0.648        & 0.201            & 0.018          & 23.24    & 224.6    & V       \\
            &                          & V    & 9  & 0.649        & 0.184            & 0.048          & 8.73     & 216.2    & V       \\
            &                          & R    & 9  & 0.592        & 0.178            & 0.059          & 6.48     & 199.7    & V       \\
            &                          & I    & 9  & 0.584        & 0.179            & 0.069          & 5.55     & 188.7    & V       \\
            &                          & B$-$V  & 9  &              &                  &                & 0.52     & 9.5      & NV      \\
            &                          & V$-$R  & 9  &              &                  &                & 0.65     & 16.6     & NV      \\
            &                          & R$-$I  & 9  &              &                  &                & 2.03     & 13.3     & NV      \\
            &                          & B$-$I  & 9  &              &                  &                & 4.10     & 38.2     & V       \\
PKS 0735+178& 2009 01 20 - 2009 04 16  & B    & 2  & 0.036        & 0.042            & 0.005          & 6.87     & 5.1      & PV       \\
            &                          & V    & 5  & 0.161        & 0.129            & 0.039          & 3.67     & 38.7     & V       \\
            &                          & R    & 7  & 0.087        & 0.083            & 0.017          & 4.97     & 24.1     & V       \\
            &                          & I    & 5  & 0.087        & 0.093            & 0.026          & 3.44     & 22.7     & V       \\
            &                          & B$-$V  & 2  &              &                  &                & 0.90     & 1.0      & NV      \\
            &                          & V$-$R  & 5  &              &                  &                & 1.91     & 20.4     & NV      \\
            &                          & R$-$I  & 5  &              &                  &                & 7.61     & 10.3     & V       \\
            &                          & B$-$I  & 2  &              &                  &                & 0.44     & 0.4      & NV      \\
OJ 287      & 2008 10 24 - 2009 05 26  & B    & 22 & 0.250        & 0.269            & 0.053          & 4.89     & 111.7    & V       \\
            &                          & V    & 24 & 0.241        & 0.240            & 0.025          & 9.62     & 110.2    & V       \\
            &                          & R    & 24 & 0.268        & 0.267            & 0.026          & 10.28    & 118.2    & V       \\
            &                          & I    & 24 & 0.274        & 0.272            & 0.027          & 10.21    & 121.5    & V       \\
            &                          & B$-$V  & 22 &              &                  &                & 1.51     & 42.3     & NV      \\
            &                          & V$-$R  & 24 &              &                  &                & 2.09     & 22.3     & NV      \\
            &                          & R$-$I  & 24 &              &                  &                & 2.19     & 16.9     & NV      \\
            &                          & B$-$I  & 22 &              &                  &                & 2.56     & 44.5     & NV      \\
4C 29.45    & 2009 03 25 - 2009 06 14  & B    & 11 & 0.195        & 0.178            & 0.030          & 6.15     & 59.7     & V       \\
            &                          & V    & 19 & 0.431        & 0.391            & 0.111          & 3.69     & 192.2    & V       \\
            &                          & R    & 25 & 0.365        & 0.370            & 0.045          & 8.11     & 109.9    & V       \\
            &                          & I    & 21 & 0.377        & 0.376            & 0.026          & 14.45    & 125.9    & V       \\
            &                          & B$-$V  & 11 &              &                  &                & 3.58     & 84.7     & V       \\
            &                          & V$-$R  & 19 &              &                  &                & 9.22     & 97.2     & V       \\
            &                          & R$-$I  & 21 &              &                  &                & 3.68     & 65.8     & V       \\
            &                          & B$-$I  & 11 &              &                  &                & 7.13     & 116.3    & V       \\
3C 273      & 2009 01 20 - 2009 06 16  & B    & 44 & 0.061        & 0.057            & 0.041          & 1.43     & 28.1     & NV      \\
            &                          & V    & 38 & 0.022        & 1.456            & 1.458          & 0.51     & 10.6     & NV      \\
            &                          & R    & 38 & 0.018        & 0.019            & 0.017          & 1.09     & 8.4      & NV      \\
            &                          & I    & 40 & 0.024        & 0.038            & 0.031          & 0.99     & 12.5     & NV      \\
            &                          & B$-$V  & 38 &              &                  &                & 0.04     & 21.9     & NV      \\
            &                          & V$-$R  & 38 &              &                  &                & 0.01     & 9.1      & NV      \\
            &                          & R$-$I  & 38 &              &                  &                & 1.27     & 11.9     & NV      \\
            &                          & B$-$I  & 38 &              &                  &                & 1.76     & 23.5     & NV      \\\hline
\end{tabular}
\end{table*}

\begin{table*}
{Table 4. continued ... }

\begin{tabular}{lclrcccccc} \hline
3C 279      & 2009 03 25 - 2009 06 15  & V    & 16 & 0.370        & 0.376            & 0.101          & 3.68     & 124.1    & V       \\
            &                          & R    & 14 & 0.353        & 0.362            & 0.046          & 7.70     & 110.0    & V       \\
            &                          & I    & 16 & 0.295        & 0.294            & 0.044          & 6.69     & 83.3     & V       \\
            &                          & V$-$R  & 14 &              &                  &                & 1.62     & 52.9     & NV      \\
            &                          & R$-$I  & 14 &              &                  &                & 3.29     & 45.7     & V       \\
PKS 1510$-$089& 2009 04 17 - 2009 06 24& B    & 7  & 0.308        & 0.319            & 0.018          & 17.56    & 85.2     & V       \\
            &                          & V    & 8  & 0.366        & 0.375            & 0.017          & 22.01    & 101.2    & V       \\
            &                          & R    & 8  & 0.397        & 0.389            & 0.014          & 29.02    & 109.1    & V       \\
            &                          & I    & 8  & 0.382        & 0.373            & 0.014          & 26.38    & 104.9    & V       \\
            &                          & B$-$V  & 7  &              &                  &                & 2.76     & 15.9     & V      \\
            &                          & V$-$R  & 8  &              &                  &                & 1.91     & 12.1     & NV      \\
            &                          & R$-$I  & 8  &              &                  &                & 8.24     & 26.9     & V      \\
            &                          & B$-$I  & 7  &              &                  &                & 5.13     & 27.9     & V       \\
BL Lac      & 2008 09 04 - 2009 04 17  & B    & 15 & 0.169        & 0.137            & 0.049          & 3.08     & 53.0     & V       \\
            &                          & V    & 15 & 0.113        & 0.107            & 0.017          & 6.25     & 31.6     & V       \\
            &                          & R    & 19 & 0.107        & 0.091            & 0.029          & 3.42     & 30.1     & V       \\
            &                          & I    & 15 & 0.118        & 0.105            & 0.024          & 4.72     & 35.0     & V       \\
            &                          & B$-$V  & 15 &              &                  &                & 2.99     & 23.4     & NV      \\
            &                          & V$-$R  & 15 &              &                  &                & 2.42     & 8.9      & NV       \\
            &                          & R$-$I  & 15 &              &                  &                & 0.99     & 4.3      & NV       \\
            &                          & B$-$I  & 15 &              &                  &                & 1.97     & 24.1     & NV      \\
3C 454.3    & 2008 09 06 - 2008 10 30  & B    & 11 & 0.240        & 0.238            & 0.009          & 25.67    & 79.9     & V       \\
            &                          & V    & 11 & 0.252        & 0.247            & 0.017          & 14.47    & 80.6     & V       \\
            &                          & R    & 13 & 0.298        & 0.289            & 0.016          & 17.84    & 95.2     & V       \\
            &                          & I    & 11 & 0.343        & 0.329            & 0.023          & 14.53    & 110.8    & V       \\
            &                          & B$-$V  & 11 &              &                  &                & 3.93     & 11.7     & V       \\
            &                          & V$-$R  & 11 &              &                  &                & 4.29     & 15.9     & V       \\
            &                          & R$-$I  & 11 &              &                  &                & 3.13     & 15.6     & V       \\
            &                          & B$-$I  & 11 &              &                  &                & 5.99     & 39.1     & V       \\\hline

\end{tabular}  \\
$ $V : Variable, NV : Non-Variable, PV : Partially Variable    \\
$*$Colour variability  is investigated using only one comparision star, as its non-variability was
established while searching for flux variations.  \\
\end{table*}

\begin{table*}
\caption{Fits to colour-magnitude dependences and Colour-magnitude correlation coefficients}

\noindent
\begin{tabular}{lrrrrrrrr} \hline
Source Name    &\multicolumn {2} {c} {B$-$V vs V}  &\multicolumn {2} {c} {V$-$R vs V}  &\multicolumn {2} {c} {R$-$I vs V} &\multicolumn {2} {c} {B$-$I vs V} \\
               &      $m^a$      & $c^a$               &     $m$      &      $c$         &  $m $       &   $c $           &   $m$      &    $c$           \\
               &    $r^a$       &$p^a$               &$r$           &$p$               & $r$        &$p$               &$r$        &$p$               \\\hline
3C 66A         & $-$0.02     & 0.79            & 0.03       &$-$0.05         & 0.03      &  0.04          & 0.04     & 0.78            \\
               &$-$0.386     & 0.2397          &  0.601     & 0.050          & 0.670     & 0.02385        & 0.509    & 0.1098          \\
AO 0235+164    & 0.007       & 0.80            & 0.01       & 0.53           & 0.08      & $-$0.50        & 0.09     & 0.92            \\
               &  0.094      & 0.7814          &  0.273     & 0.4438         & 0.302     & 0.3955         & 0.355    & 0.2836          \\
PKS 0420$-$014 & $-$0.61     & 11.19           & $-$0.42    & 7.77           & $-$0.22   & 4.46           & $-$1.25  & 23.43           \\
               & $-$0.818    & 0.0462          & $-$0.835   & 0.0383         &$-$0.474   & 0.3414         &$-$0.901  & 0.0141          \\
S5 0716+714    & 0.03        & 0.06            & 0.06       & $-$0.44        & 0.04      & $-$0.07        & 0.13     & $-$0.44         \\
               &  0.655      & 0.0079          &  0.814     & 0.0002         & 0.729     & 0.0020         & 0.850    & 5.9e-05         \\
PKS 0735+178   & *           & *               & 0.42       & $-$6.37        & 0.16      & $-$2.06        &*         & *               \\
               & *           & *               &  0.805     & 0.0999         & 0.415     & 0.5846         &*         & *               \\
OJ 287         & $-$0.09     & 4.21            & $-$0.06    & 1.34           & $-$0.02   & 0.98           & $-$0.17  & 4.21            \\
               &$-$0.437     & 0.0475          & $-$0.249   & 0.1840         &$-$0.091   & 0.6309         &$-$0.329  & 0.1342          \\
4C 29.45       & $-$0.63     & 11.71           & $-$0.13    & 2.47           & 0.03      & $-$0.002       & $-$0.73  & 14.45           \\
               & $-$0.665    & 0.050           & $-$0.235   & 0.3175         & 0.059     & 0.803          &$-$0.504  & 0.1665          \\
3C 273         & $-$0.78     & 10.16           & 0.62       & $-$7.76        & $-$0.41   & 5.68           & $-$5.67  & 8.08            \\
               & $-$0.258    & 0.0905          &  0.650     & 1.3e-06        &$-$0.413   &0.0048          &$-$0.176  & 0.2523          \\
3C 279         & *           & *               & 0.22       & $-$3.32        & 0.15      & $-$1.82        & *        & *               \\
               & *           & *               &  0.603     & 0.0134         & 0.585     & 0.0277         & *        & *               \\
PKS 1510$-$089 & $-$0.14     & 2.69            &$-$0.08     & 1.78           & 0.07      & $-$0.64        & $-$0.08  & 2.71            \\
               & $-$0.952    & 0.0009          & $-$0.685   & 0.0606         & 0.305     & 0.4613         &$-$0.264  & 0.5662          \\
BL Lac         & 0.44        & $-$5.46         & 0.003      & 0.722          & 0.01      & 0.66           & 0.42     & $-$3.58         \\
               &  0.713      & 0.0091          &  0.014     & 0.9661         & 0.109     & 0.7485         & 0.639    & 0.0253          \\
3C 454.3       & $-$0.06     & 1.44            & $-$0.18    & 3.27           & $-$0.17   & 3.22           & $-$0.40  & 7.94            \\
               & $-$0.378    & 0.2516          & $-$0.869   &0.0005          &$-$0.79    &0.0033          &$-$0.837  &0.0013           \\\hline
\end{tabular}  \\
$^a$ $m =$ slope and $c =$ intercept of CI against V; $r =$ Pearson coefficient, $p =$ null hypothesis probability\\
$*$ missing entry is due to lack of data \\
\end{table*}
\clearpage

\clearpage

\begin{figure*}
\epsfig{figure = 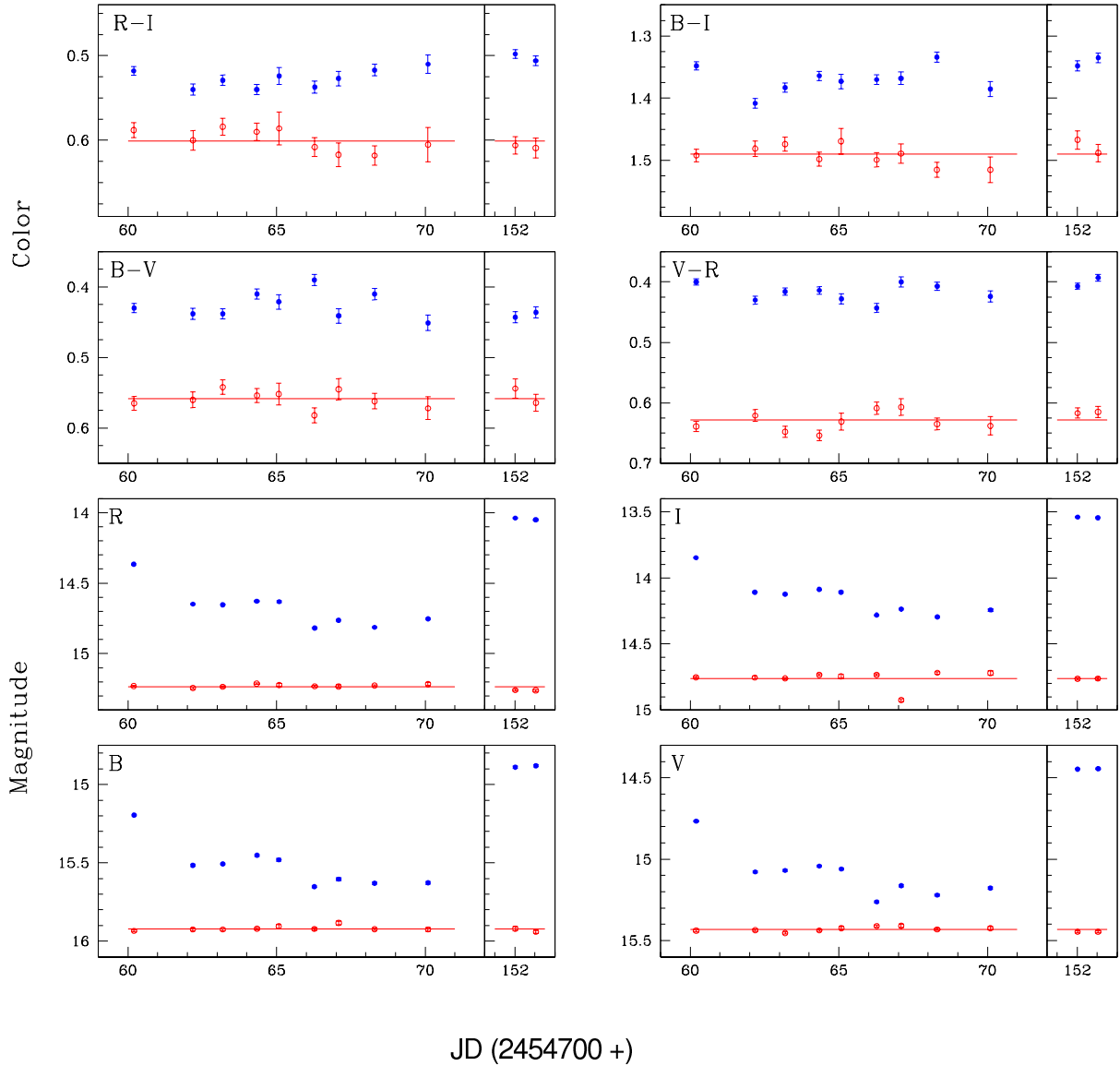,height=16.cm,width=16.cm,angle=0}
\caption{Calibrated LCs of 3C 66A (w.r.t. star 2) are in the lower four panels, plotted with the differential instrumental 
magnitudes of stars 2 and 3 with arbitrary offsets.  The colour LCs in the upper four panels are 
plotted with the corresponding 
colours of star 2.  Empty circles are the standard stars' LCs with arbitrary offsets and the straight line represents the mean of the standard stars' LCs.}     

\end{figure*}

\begin{figure*}
\epsfig{figure= 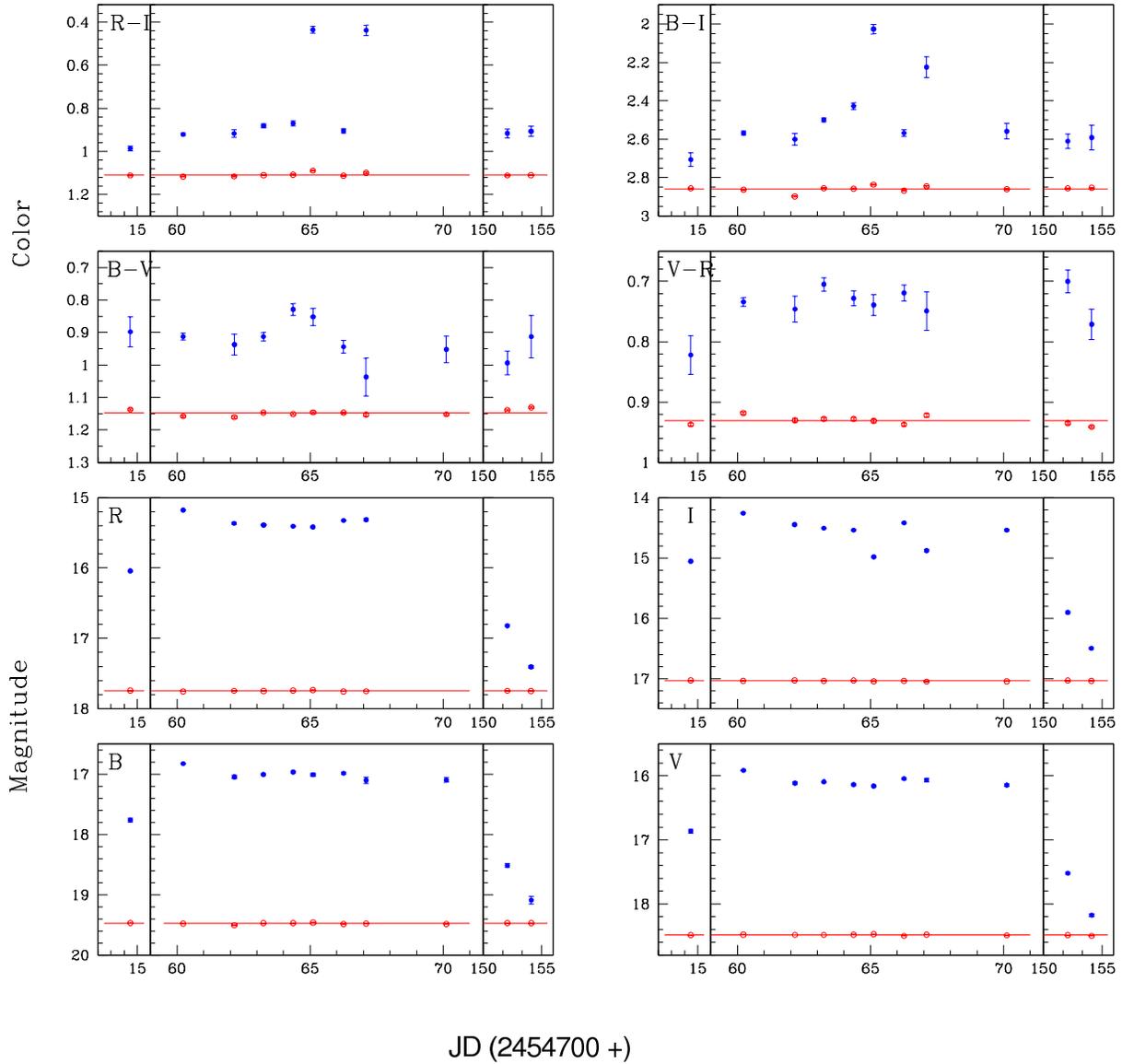,height=16.cm,width=16.cm,angle=0}
\caption{As in Fig.\ 1 for AO 0235$+$164. The comparison stars are 2 and 3; 2 is the calibrator.} 
\end{figure*}

\begin{figure*}
\epsfig{figure= 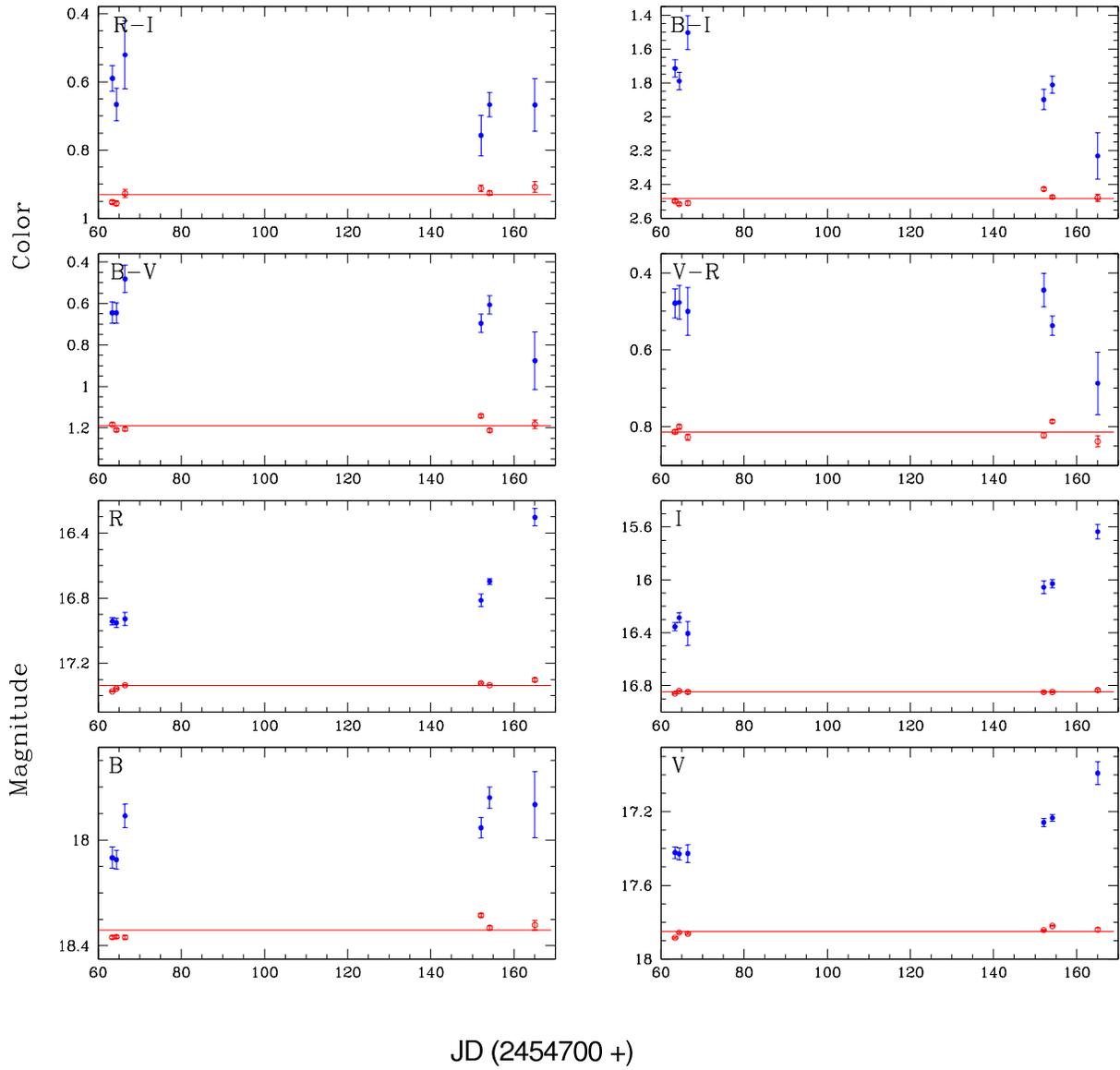,height=16.cm,width=16.cm,angle=0}
\caption{As in Fig.\ 1 for  PKS 0420$-$014. The comparison stars are 4 and 2;  4 is the calibrator.} 
\end{figure*}

\begin{figure*}
\epsfig{figure= 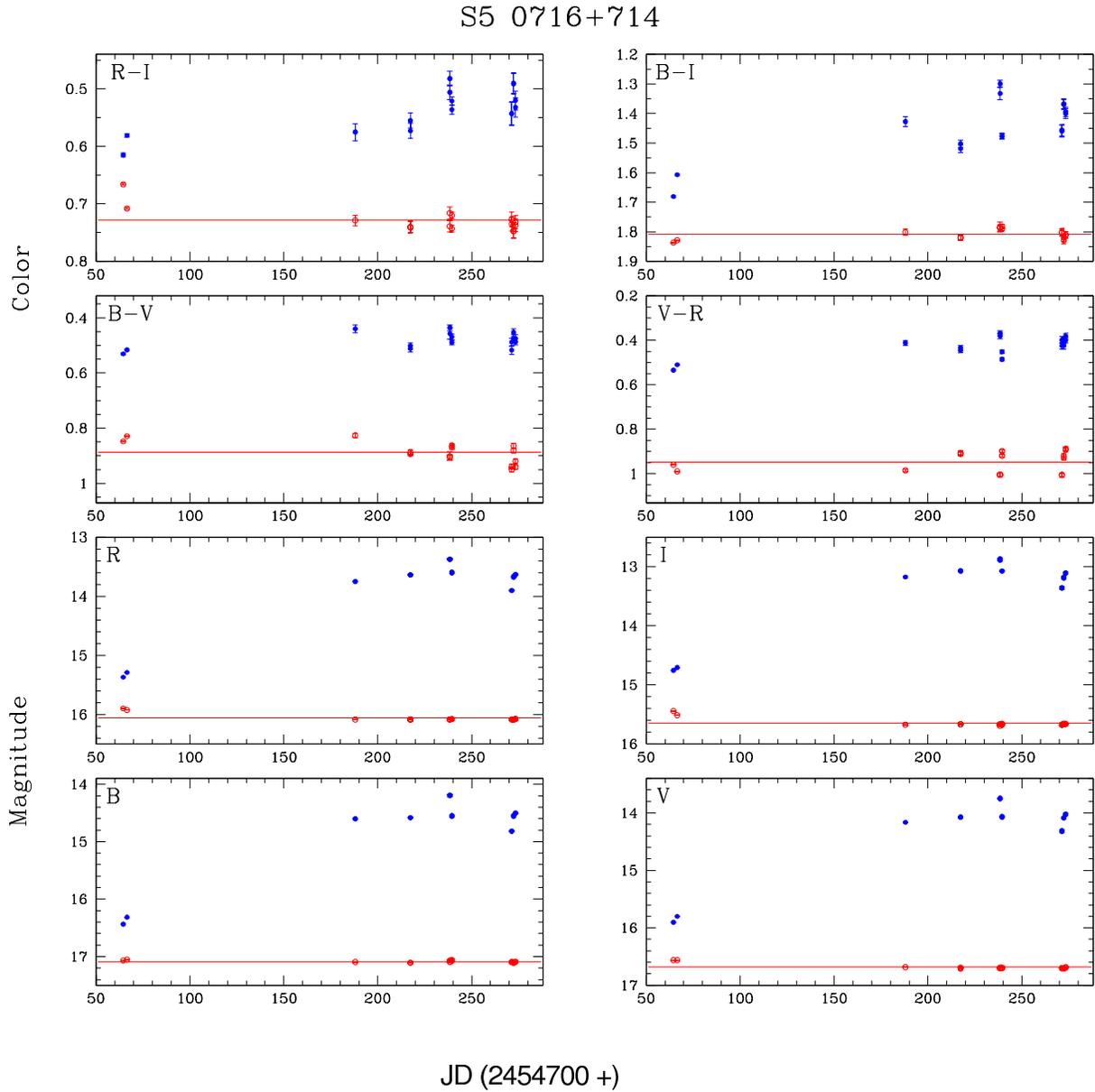,height=16.cm,width=16.cm,angle=0}
\caption{As in Fig.\ 1 for S5 0716$+$714. The comparison stars are 5 and 3; 5 is the calibrator.} 
\end{figure*}

\begin{figure*}
\epsfig{figure= 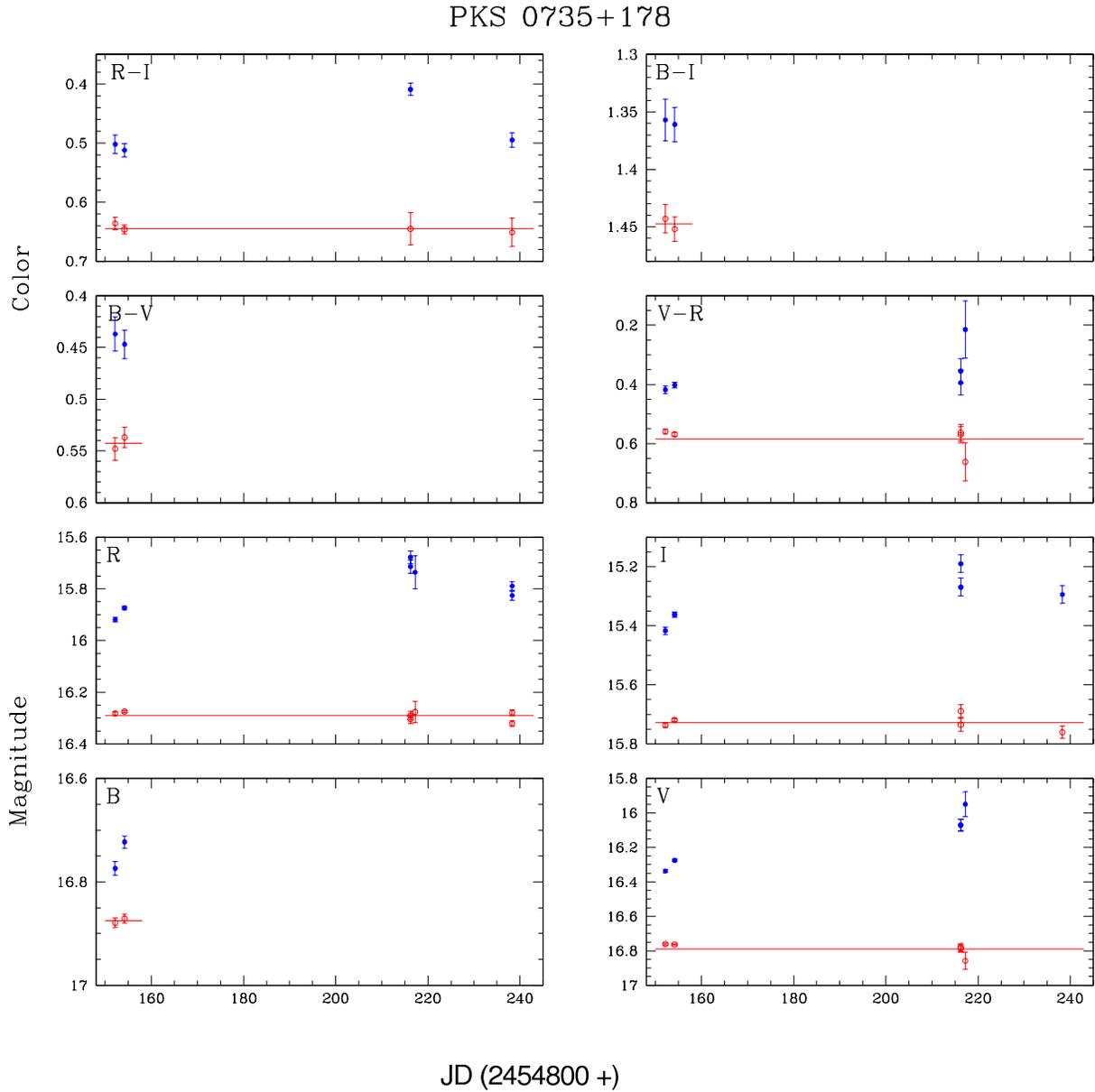,height=16.cm,width=16.cm,angle=0}
\caption{As in Fig.\ 1 for  PKS 0735$+$178. The comparison stars are D and C; D is the calibrator.}
\end{figure*}

\begin{figure*}
\epsfig{figure= 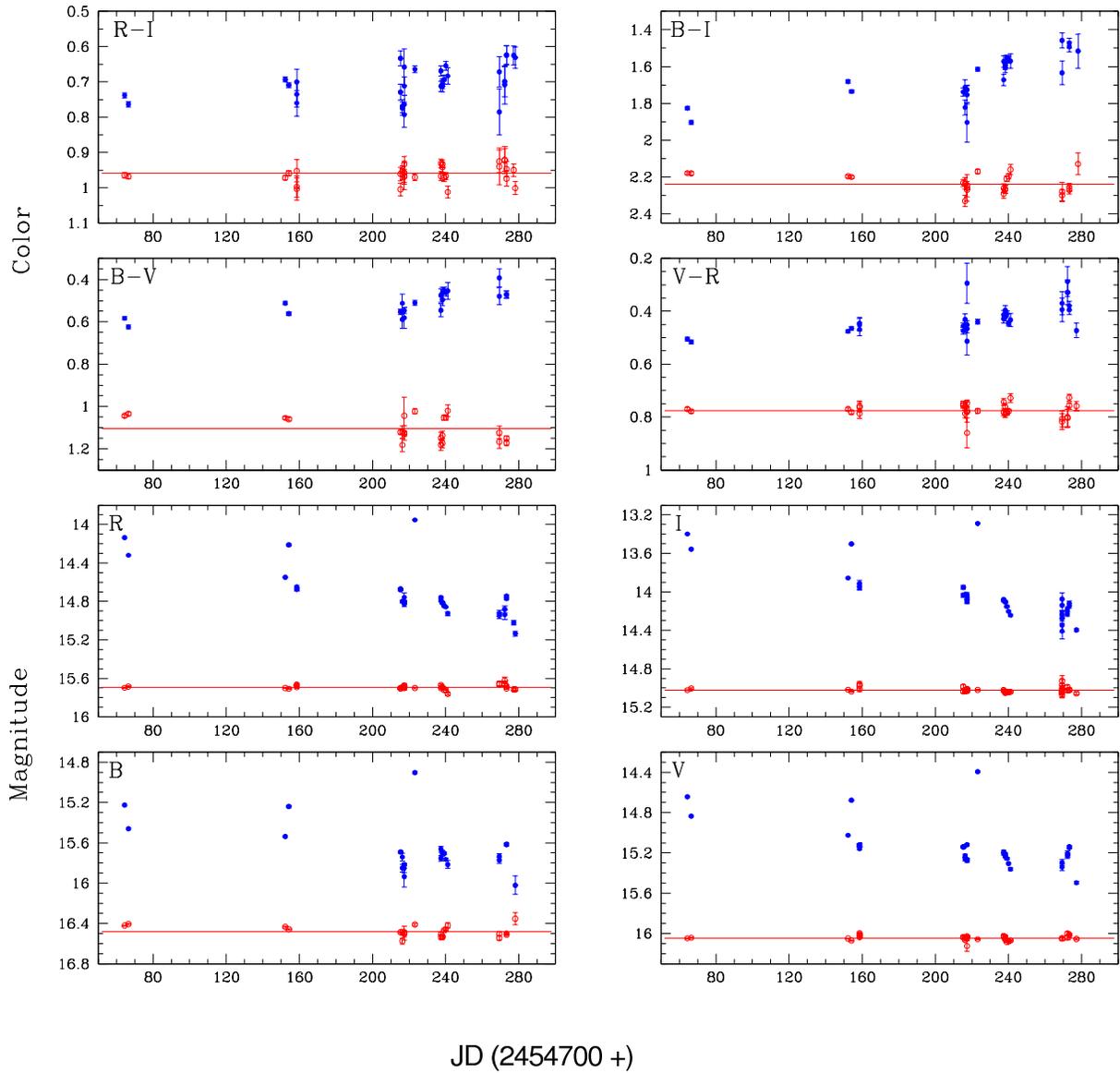,height=16.cm,width=16.cm,angle=0}
\caption{As in Fig.\ 1 for  OJ 287. The comparison stars are  11 and 4; star 11 is the calibrator.}
\end{figure*}

\begin{figure*}
\epsfig{figure= 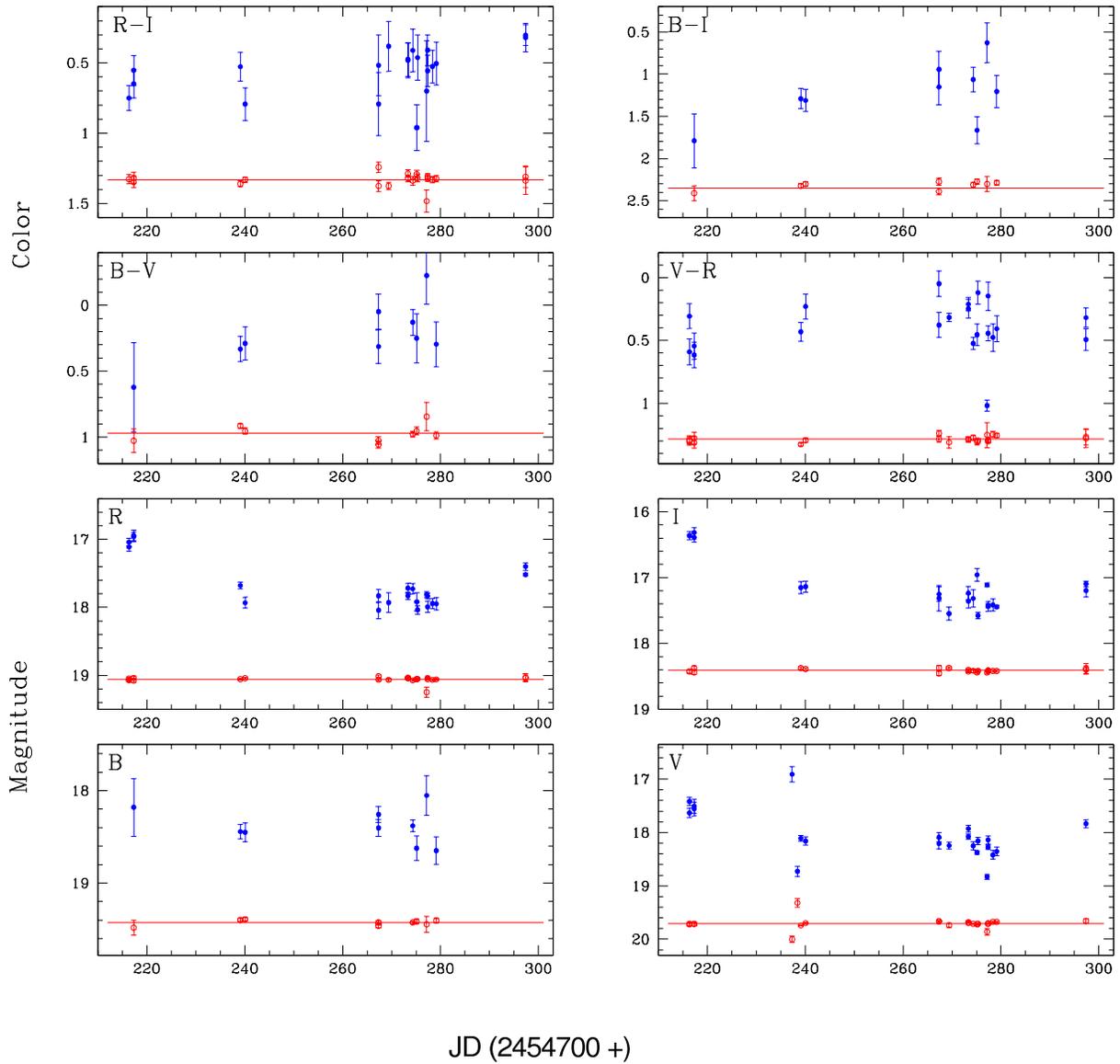,height=16.cm,width=16.cm,angle=0}
\caption{As in Fig.\ 1 for 4C 29.45. The comparison stars are 14 and 13; 14 is the calibrator.} 
\end{figure*}

\begin{figure*}
\epsfig{figure= 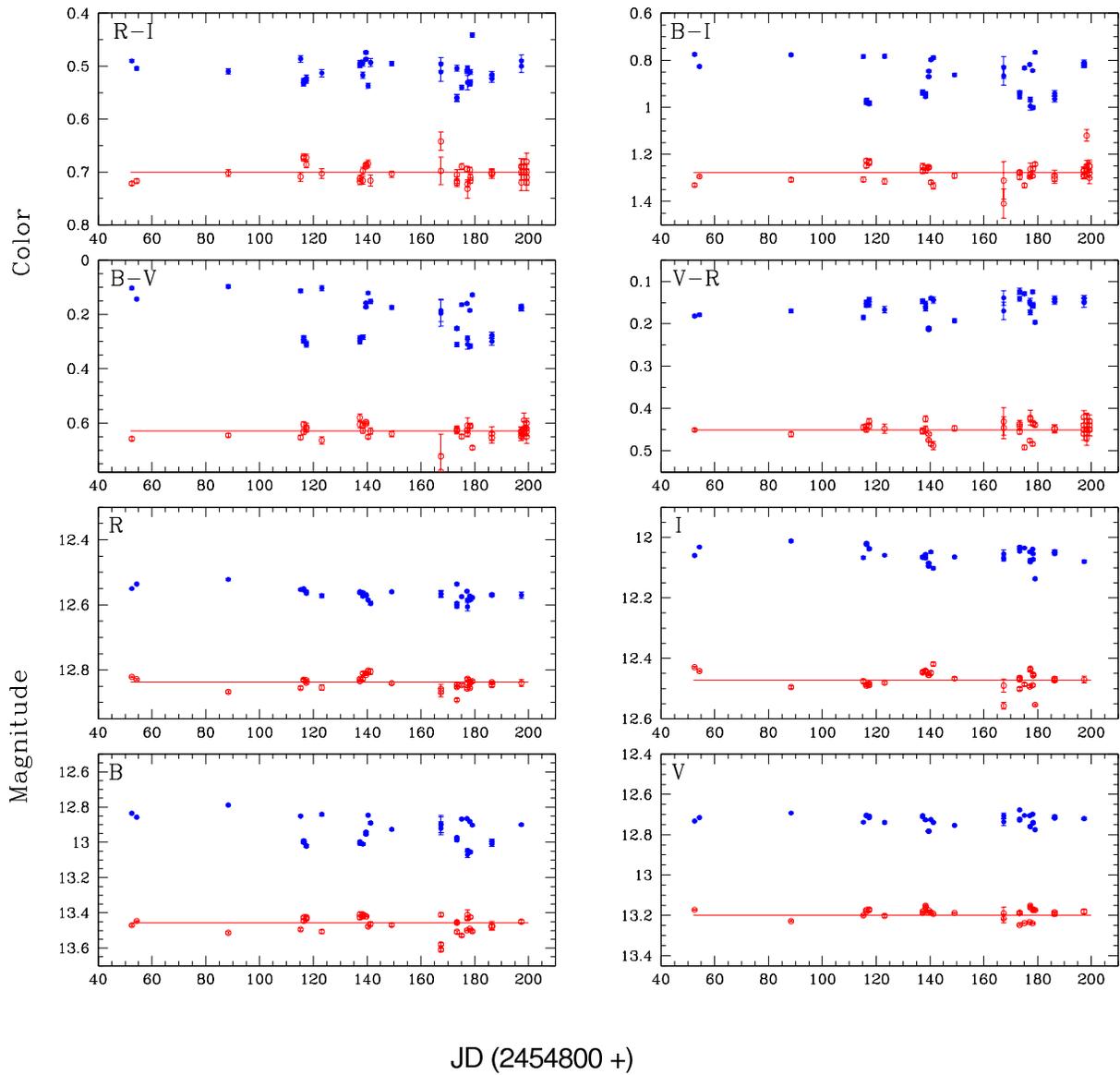,height=16.cm,width=16.cm,angle=0}
\caption{As in Fig.\ 1 for 3C 273. The comparison stars are E and G; E is the calibrator.}
\end{figure*}

\begin{figure*}
\epsfig{figure= 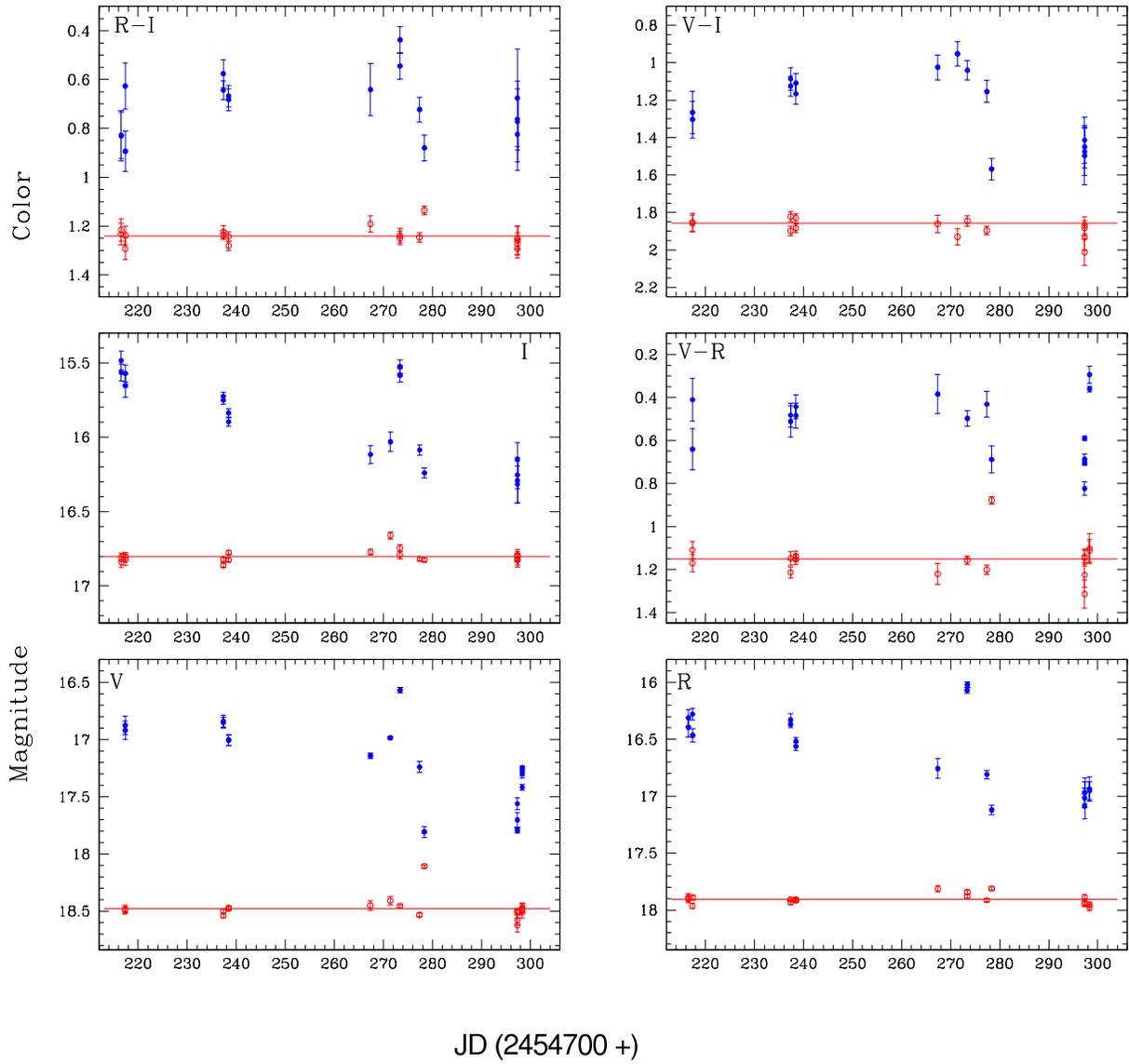,height=16.cm,width=16.cm,angle=0}
\caption{As in Fig.\ 1 for 3C 279. The comparison stars are  1 and 4; 1 is the calibrator.}
\end{figure*}

\begin{figure*}
\epsfig{figure= 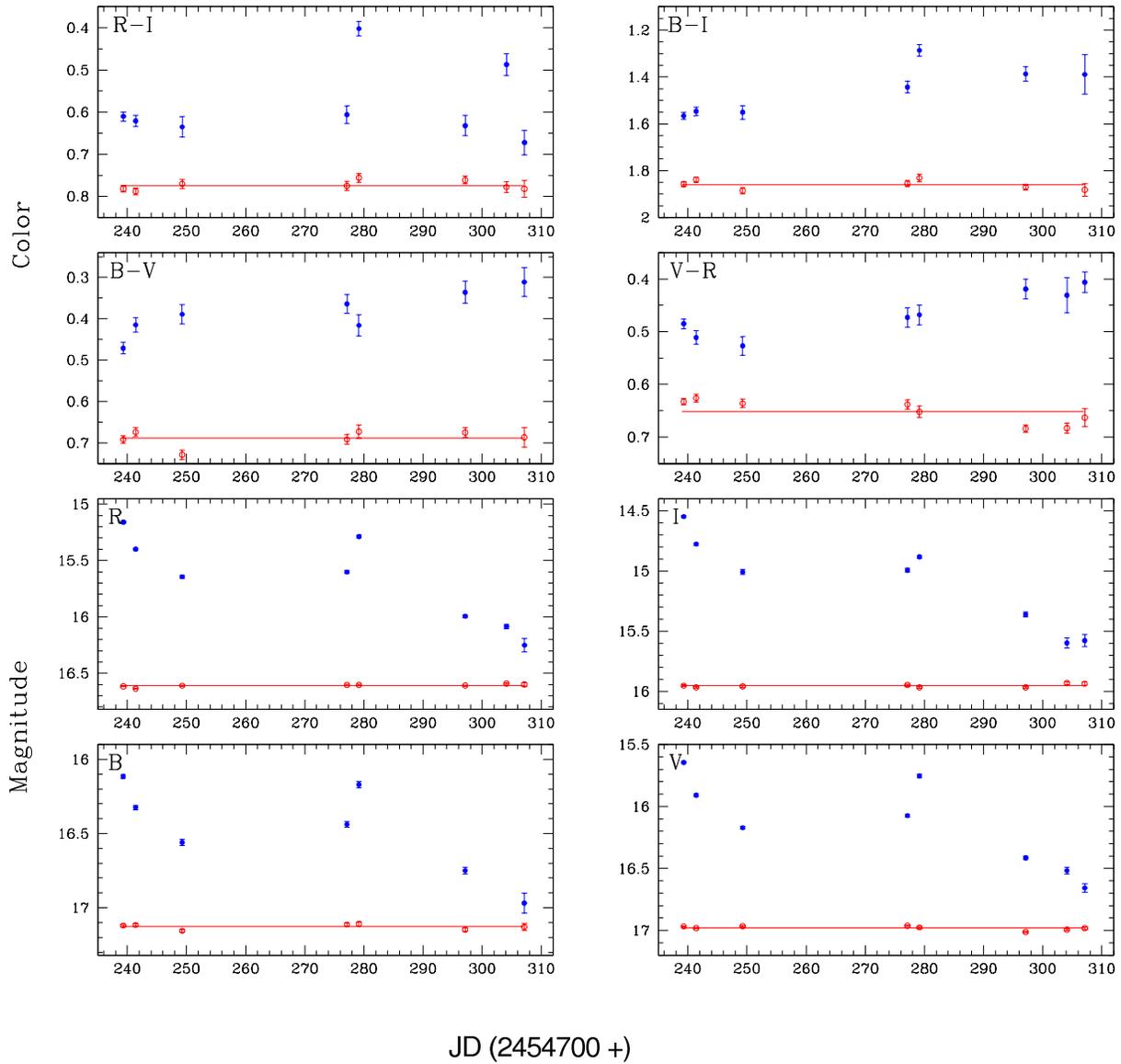,height=16.cm,width=16.cm,angle=0}
\caption{As in Fig.\ 1 for PKS 1510$-$089. The comparison stars are  3 and 4; 3 is the calibrator.}
\end{figure*}

\begin{figure*}
\epsfig{figure= 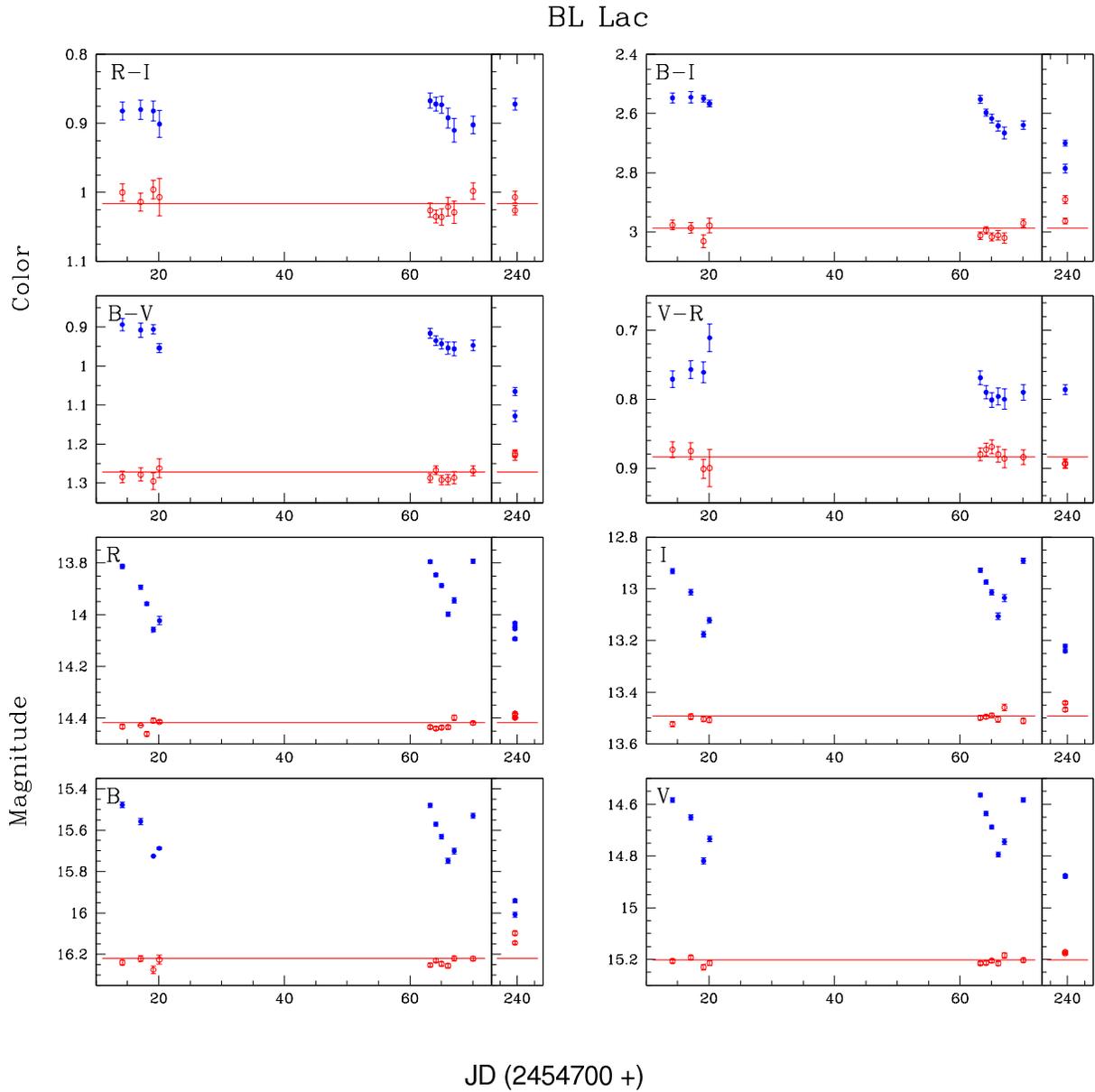,height=16.cm,width=16.cm,angle=0}
\caption{As in Fig.\ 1 for BL Lac. The comparison stars are  K and C;  K is the calibrator.} 
\end{figure*}

\begin{figure*}
\epsfig{figure= 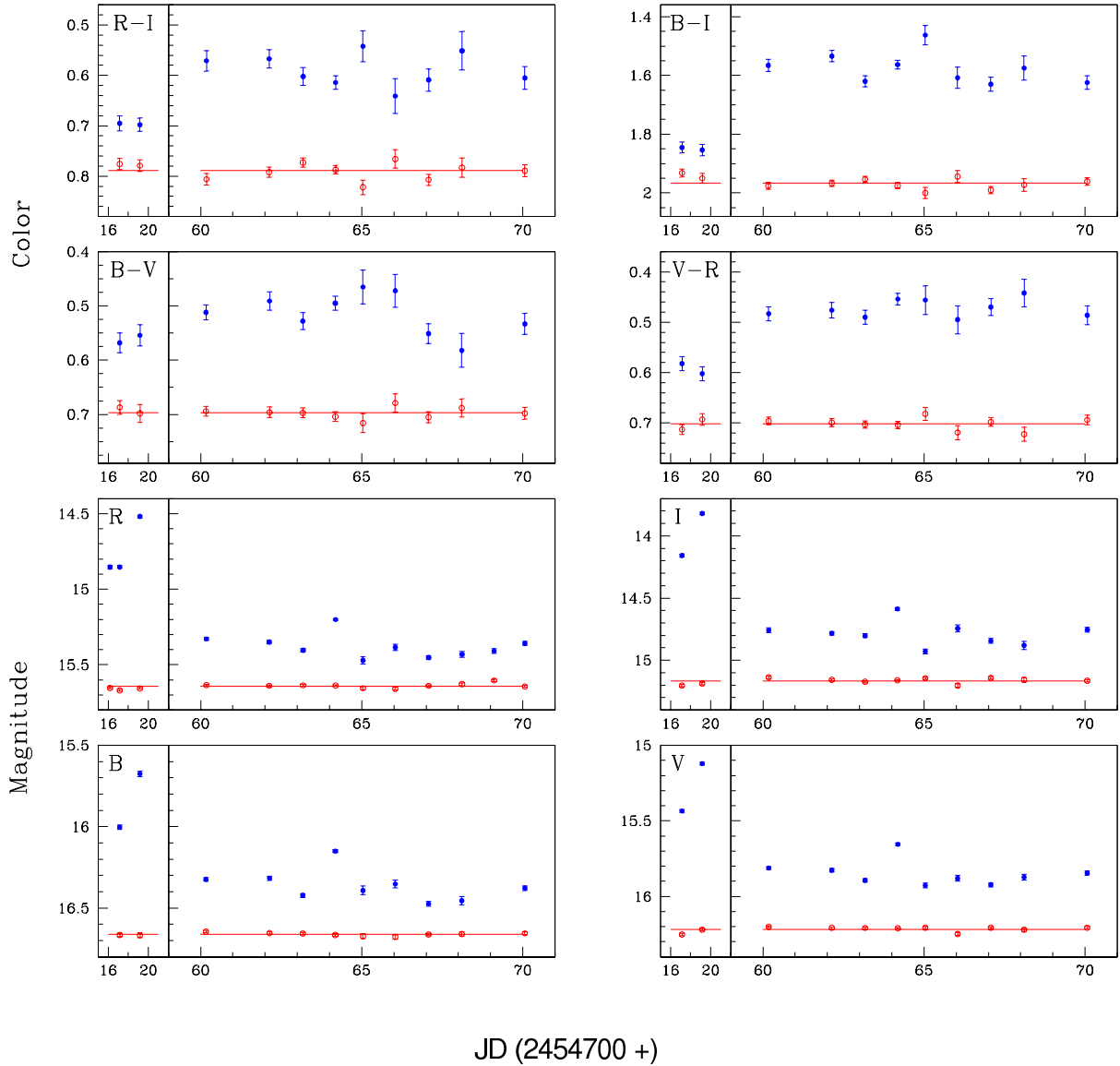,height=16.cm,width=16.cm,angle=0}
\caption{As in Fig.\ 1 for 3C 454.3. The comparison stars are  B and C; B is the calibrator.} 
\end{figure*}

\begin{figure*}
\epsfig{figure= 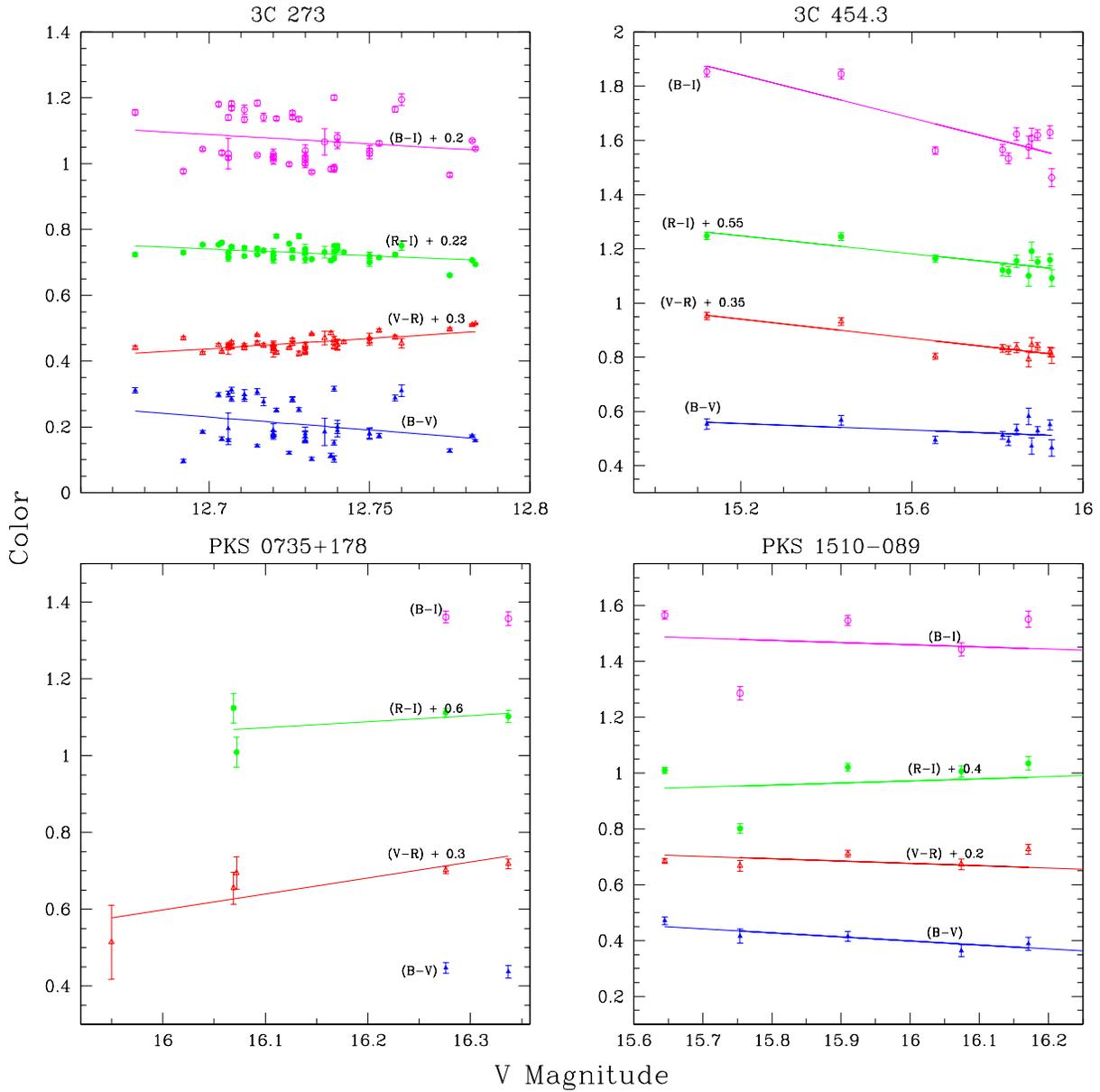,height=16.cm,width=16.cm,angle=0}
\caption{Colour--magnitude plots of the blazars 3C 273, 3C 454.3, PKS 0735+178 \& PKS 1510-089.}
\end{figure*}

\begin{figure*}
\epsfig{figure= 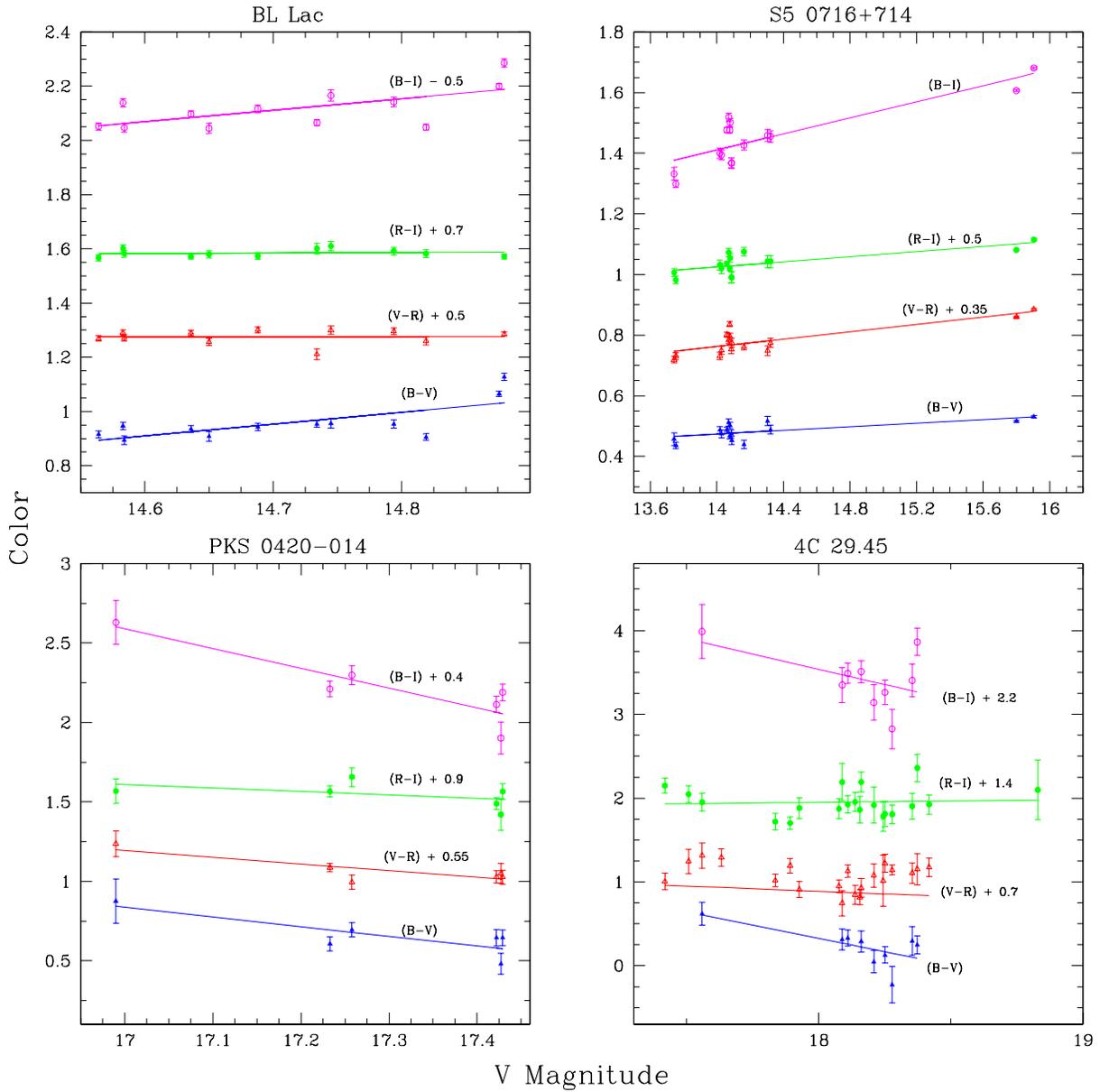,height=16.cm,width=16.cm,angle=0}
\caption{Colour--magnitude plots of the blazars BL Lac, S5 0716+714, PKS 0420-014 \& 4C 29.45.}
\end{figure*}

\begin{figure*}
\epsfig{figure= 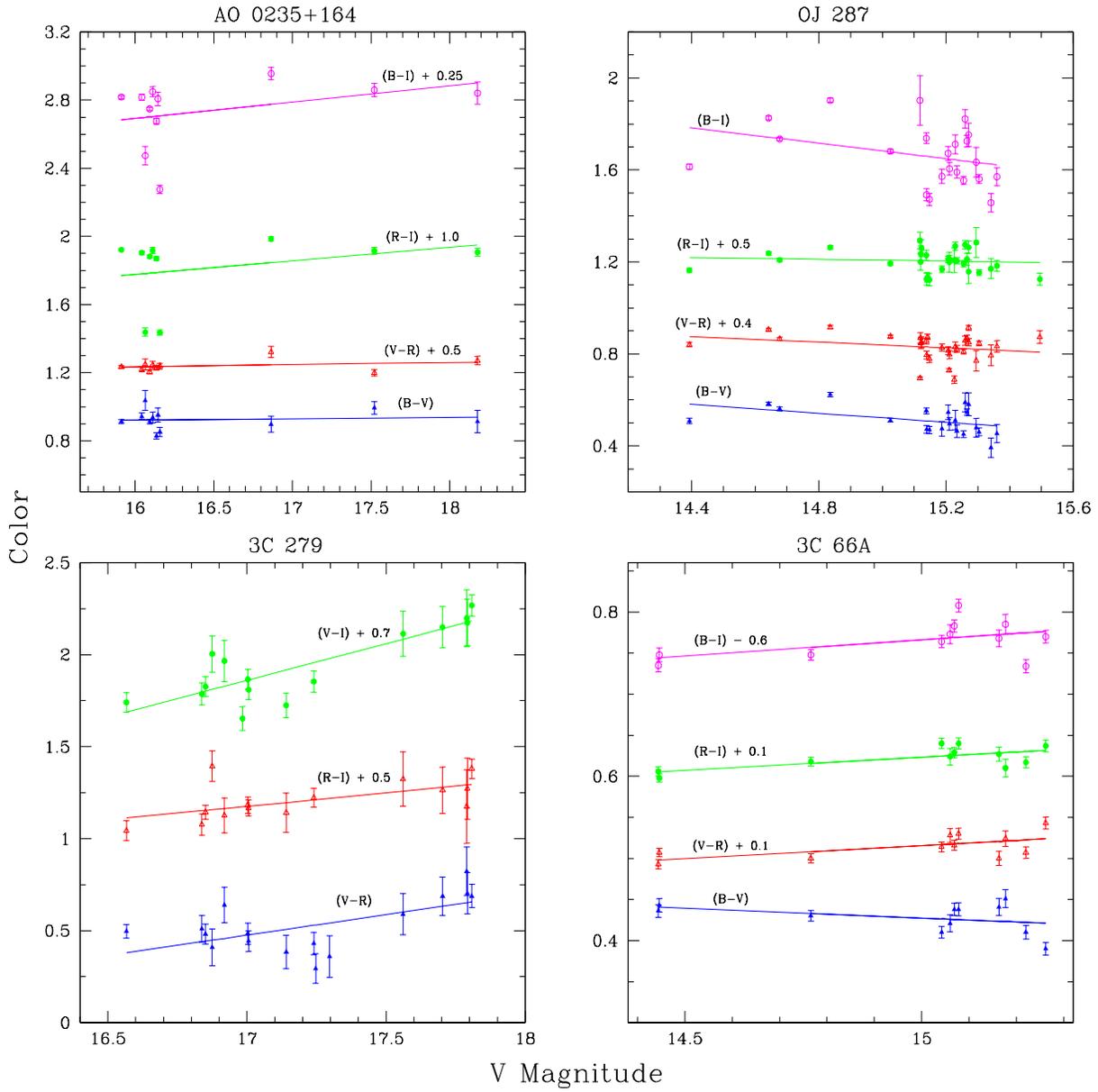,height=16.cm,width=16.cm,angle=0}
\caption{Colour--magnitude plots of the blazars AO 0235+164, OJ 287, 3C 279,  \& 3C 66A.}
\end{figure*}

\end{document}